\documentclass[11pt]{article}
\usepackage{jheppub} 
\usepackage{amsmath,amssymb}
\usepackage[usenames,dvipsnames]{xcolor}
\usepackage{subcaption}

\newcommand{\ket}[1]{|#1\rangle}
\newcommand{\bra}[1]{\langle #1|}
\newcommand{\braket}[2]{\langle #1|#2 \rangle}
\newcommand{\vev}[1]{\langle #1 \rangle}

\newcommand\e{\text{e}}
\newcommand\Tr{\mathrm{Tr}}

\hypersetup{
    pdfencoding=unicode,
	colorlinks=true,
	urlcolor=Maroon,
	linkcolor=RoyalBlue,
	citecolor=Maroon,
	pdftitle={Goldilocks bootstrap}, 				
	pdfauthor={David Berenstein, Victor A. Rodriguez},
	pdfdisplaydoctitle=true,
	pdfstartview=FitH,
	linktocpage=true
}

\definecolor{bleudefrance}{rgb}{0.19, 0.55, 0.91}
\definecolor{candyapplered}{rgb}{1.0, 0.03, 0.0}

\title{Goldilocks and the bootstrap}

\author{David Berenstein$^{\dagger,\ddagger,\triangle}$ and Victor A. Rodriguez$^\dagger$}
\affiliation{$^\dagger$Department of Physics, University of California, Santa Barbara, CA 93106, USA}
\affiliation{$^\ddagger$Institute of Physics, University of Amsterdam, Science Park 904, PO Box 94485, 1090 GL Amsterdam,
The Netherlands}
\affiliation{$^\triangle$Delta Institute for Theoretical Physics, Science Park 904, PO Box 94485, 1090 GL Amsterdam, The
Netherlands}

\emailAdd{dberens@physics.ucsb.edu, varodriguez@ucsb.edu}

\abstract{We study simplified bootstrap problems for probability distributions on the infinite line and the circle. We show that the rapid convergence of the bootstrap method for problems on the infinite line is related to the fact that the smallest eigenvalue of the positive matrices in the exact solution becomes exponentially small for large matrices, while the moments grow factorially. As a result, the positivity condition is very finely tuned.
For problems on the circle we show instead that the entries of the positive matrix of Fourier modes of the distribution depend linearly on the initial data of the recursion, with factorially growing coefficients. By positivity, these matrix elements are bounded in absolute value by one, so the initial data must also be fine-tuned. Additionally, we find that we can largely bypass the semi-definite program (SDP) nature of the problem on a circle by recognizing that these Fourier modes must be asymptotically exponentially small. 
With a simple ansatz, which we call the shoestring bootstrap, we can efficiently identify an interior point of the set of allowed matrices with much higher precision than conventional SDP bounds permit. 
We apply this method to solving unitary matrix model integrals by numerically constructing the orthogonal polynomials associated with the circle distribution.}

\begin{document}
\maketitle
\section{Introduction}
\label{sec:intro}

The original idea of the bootstrap was that analyticity, unitarity and crossing symmetry of the S-matrix was enough to determine physical theories that describe particle physics. Our modern understanding of this idea is that there is no unique solution of the bootstrap program. Field theories like the Standard Model of particle physics each produce a solution of the bootstrap program, where the masses and interactions of the particles appearing in the S-matrix are determined from the parameters of UV theory. In that sense, the bootstrap becomes a set of consistency conditions that can be used to obtain bounds on physical phenomena when other techniques are not necessarily enough to compute the full answer we seek. 

On the other hand, the bootstrap program has been able to solve a wide variety of problems. These require additional input. Starting with the analytic solution of conformal field theories (CFT) in $d=2$ dimensions \cite{Belavin:1984vu}, critical exponents were determined exactly by finding that there was an infinite symmetry underlying these conformal field theories.
Unitarity of the representation theory of the conformal group was enough to solve the theories in cases where the central charge satisfies $c<1$. This method produces the list of the minimal models.
More recently, conformal field theories in higher dimensions have been successfully tackled from this point of view \cite{Poland:2018epd}. Starting with the works \cite{El-Showk:2012cjh,El-Showk:2014dwa} it was shown that precise numerical bounds for critical exponents and other CFT data could be obtained for the three dimensional Ising model. Here, the bootstrap program is for a Euclidean field theory. Unitarity gets replaced by reflection positivity and crossing symmetry is related to the associativity of the OPE expansion, rather than the analyticity of the scattering S-matrix. 

In recent years, bootstrap techniques have significantly broadened their impact within the physics and applied mathematics communities, with applications ranging from matrix integrals \cite{Anderson:2016rcw,Lin:2020mme,Kazakov:2021lel}, lattice quantum and statistical field theory \cite{Kazakov:2022xuh,Cho:2022lcj,Berenstein:2024ebf,Li:2024wrd,Kazakov:2024ool,Guo:2025fii}, and classical dynamical systems \cite{doi:10.1137/15M1053347}, to quantum mechanical systems \cite{Berenstein:2021dyf,Bhattacharya:2021btd,Berenstein:2021loy,Li:2022prn,Berenstein:2022unr,Berenstein:2022ygg,Nancarrow:2022wdr,Berenstein:2023ppj}, quantum many-body systems\footnote{Applications in quantum chemistry date back to \cite{PhysRevA.57.4219,10.1063/1.1360199}.} \cite{Fawzi2024,PhysRevX.14.021008,Cho:2024owx}, and matrix quantum mechanics \cite{Han:2020bkb,Lin:2023owt,Cho:2024kxn,Lin:2024vvg}, among others. 
These require a reinterpretation of the ingredients above to suit particular problems. The notion of unitarity is replaced by positive definiteness of some data. 
The fact that we need some dynamics is introduced in terms of some exact constraints or recursion between parts of the associated data. This replaces analyticity. Crossing has no natural analog for the more general problems, but we interpret it broadly as being subsumed in the dynamical input that is used to constrain the problem.

One such approach to quantum mechanical problems is the method discussed in \cite{Han:2020bkb}, which builds on ideas from \cite{Lin:2020mme} for solving matrix integrals. 
These have been subsequently explored for the real line in \cite{Berenstein:2021dyf,Bhattacharya:2021btd} and for problems on the circle in \cite{Aikawa:2021eai,Berenstein:2021loy,Tchoumakov:2021mnh}. The constraints result from assuming an eigenstate of the Hamiltonian with energy $E$ and consistency conditions that arise from dynamics
\begin{equation}
\bra{E} {\cal O} \hat H \ket E=\bra{E}  \hat H {\cal O}\ket E= E \bra{E}  {\cal O}\ket E ~,
\end{equation}
and a positivity condition,
\begin{equation}
    \bra E {\cal O}^\dagger {\cal O}\ket E \geq 0 ~.
\end{equation}
The first set of constraints can be solved into a recursion relation for moments of the position operator and  positivity of the moment problem is usually enough to solve for $E$ to high accuracy (the consistent set of $E$ shrinks exponentially fast in the size of the matrix constraints that are imposed). 
For problems on the circle one uses the Fourier modes of the probability distribution instead and one finds the band structure of the potential. This was extended to problems on the positive real axis in \cite{Berenstein:2022ygg}, where an integration by parts anomaly was necessary to initialize the recursion correctly. 
Problems on the interval can also be studied \cite{Sword:2024gvv} and even some scattering problems can be analyzed this way \cite{Berenstein:2023ppj}. 

Finding the correct values by guessing seems to amount to a finely tuned problem and just like in Goldilocks, it needs to be {\em just right}. 
The original algorithm of \cite{Han:2020bkb} generically requires searching (guessing) in a high-dimensional space. This was improved to an optimization problem at fixed energy \cite{Berenstein:2022unr}, which allows one to study fairly general Hamiltonians with an arbitrary polynomial potential and obtain good data for energy levels. 

In all these examples, it was observed empirically that the bootstrap method seems to converge exponentially fast. What is the mathematical property of the problems that are being studied that guarantees the fast convergence? 
Understanding this issue is not just an exercise ensuring that the mathematical framework is sound. 
A refined understanding could also lead to improved numerical algorithms, potentially rendering previously intractable problems manageable.
Such insights might help identify strategies to reduce the computational effort and resources needed to calculate accurate approximations to the solution. 

In this paper we study this question in a simplified setting. Rather than studying quantum mechanics, we consider instead a simpler problem with similar characteristics associated to  parametrized probability distributions on the real axis and on the circle. The advantage of these toy model problems is that the recursion equations are simplified.
The positivity constraints are the same that have been used in one dimensional quantum mechanics, so the problem of convergence to the solution should be similar, without also having to search in the space of allowed energies $E$. 
The distributions we consider are of the form
\begin{equation}
    d\mu \propto \exp( -V(x)), \quad d\mu \propto \exp( -V(\theta)) \label{eq:measure}
\end{equation}
where $V(x)$ is a polynomial and $V(\theta)$ is a real periodic function and has only a finite number of non-zero Fourier coefficients
(it is a Laurent polynomial in $z=\exp(i\theta)$ with real coefficients which is real if $\theta$ is real, or alternatively, it is a polynomial in $\cos(\theta) $ with real coefficients). For problems on the real line, we require that the distribution vanishes at infinity, with $\lim_{x\to \pm \infty} V(x)=\infty$. 

These simplified problems have applications to the study of matrix models. The idea is to use these distributions to construct the orthogonal polynomials associated to the distributions. From the orthogonal polynomials one can obtain numerical results for matrix models at finite $N$. It turns out that the problem of building these orthogonal polynomials is numerically unstable \cite{gautschi1982generating,gautschi1985orthogonal}, so very high precision data is required to generate them. More general techniques can be found in \cite{gautschi2004orthogonal}.
This is where the fact that the bootstrap problem converges quickly can produce the high precision data that is required to generate this list of polynomials sufficiently accurately. This is less of an issue for problems on the circle, except that we will show that numerical instabilities also arise when we consider large $N$ for different reasons than for problems on the line.

Of particular interest to us is to compare to the bootstrap approach of Anderson and Kruczenski to lattice quantum field theory \cite{Anderson:2016rcw} (for more recent results in this direction see \cite{Kazakov:2021lel,Kazakov:2022xuh,Kazakov:2024ool}), where the loop equations of a specific lattice problem (usually at large $N$) and positivity are used to solve matrix models.
This is different than passing to collective coordinates at large $N$ in matrix quantum mechanics \cite{Jevicki:1982jj,Jevicki:1983wu}, where the positivity resides in the kinetic term of the collective coordinates (see \cite{Mathaba:2023non} for recent numerical results in these setups). 
Our results solve  finite $N$ problems related to the ones studied in \cite{Anderson:2016rcw} and we can use these to compare the physics at finite $N$ versus infinite $N$. For example, we look at the Gross-Witten transition \cite{Gross:1980he} as a test case of our numerics.

The paper is organized as follows. In section \ref{sec:examples} we introduce the two simple problems with probability distributions on the infinite line and the circle and we solve them by bootstrap methods. We compare with the exact analytic answers and show numerically that the bootstrap problem approximates the answer with exponential accuracy in the size of the positive matrices. In section \ref{sec:real_line} we analyze the moment problem for the real line more carefully. We use known results about the smallness of eigenvalues of the moment matrices to argue that the positivity constraints is close to being violated in the exact solution. The factorial dependence on the initial data of the recursion suggests that the problem becomes very fine-tuned and it is easy to violate the positivity constraint. We also analyze a slightly improved bootstrap method that maximizes the minimal eigenvalue of the moment matrices to obtain an interior point of the SDP problem and argue that this is an effective way to find an answer, which can also be accompanied by a certificate of positivity. In section \ref{sec:circle} we study the problems on the circle. 
We still find exponential sensitivity to the initial recursion data for the large Fourier modes of the distribution $a_k$ and discuss a cheap bootstrap method that uses the simple bounds $|a_k|<1$ for large $k$ to obtain regions of exclusion for the values of $a_1$ that shrink exponentially fast. We show that this is substantially faster than the original bootstrap formulation, with similar levels of precision.
We then also input the additional information that in all these problems on the circle the Fourier modes $a_k\to 0$ exponentially fast when $k$ becomes large. We show that assuming $a_k=0$ at some large $k$ (or more generally a collection of $s$ consecutive $a_k$ when there are $s$ search parameters) provides a much better solution of the problem than the cheap bootstrap, with almost twice as many digits of precision. We label this method the shoestring bootstrap. It requires solving one problem in linear algebra with high precision. We can also obtain a certificate of validity that the point obtained this way is in the window of positivity of the matrix with only one check. 
In section \ref{sec:matrix} we apply this shoestring method to solving unitary matrix models associated to one $N\times N$ matrix. We do this by constructing numerically the orthogonal polynomials of the measure from the data of Fourier modes of the distribution that has been obtained to high accuracy. We show that at large $N$, because of the large $N$ double scaling,  there are small denominators that appear in the method that makes this process numerically unstable.
The instability justifies having very high precision for the initialization data.
We show good agreement with the expected answers for the toy problem of the Gross-Witten transition. We also compare to other methods used to study this problem. In section \ref{sec:conclusion} we conclude with a summary of our findings.

\section{Two bootstrap examples from single variable integrals}\label{sec:examples}

In this section, we introduce two simple toy models that will guide our understanding of the fast convergence of the bootstrap method for bounding moments of a given distribution. 
Particularly, the toy model on the circle will serve as a precise testing ground for our proposed cheap bootstrap algorithm described in the next section.

\paragraph{Toy example on the real line.}
Let us consider first solving a well-known toy problem by bootstrap methods that is similar to the quantum mechanical bootstrap. The idea is to consider the following integral as a probability measure
\begin{equation}
    \int dx\, \exp( -b x^2/2 -x^4/4) ~,
\end{equation}
where we think of the integral as the integral of a probability density $d\mu \propto \exp( -b x^2/2 -x^4/4) dx$. For the integral to be well-defined, we require that $b$ is real.

We can characterize the distribution by its moments, which capture important statistical properties of $d\mu$ that can be measured. These moments are defined as
\begin{equation}
    a_n = N\int_{-\infty}^\infty dx\, x^n \exp( -b x^2/2 -x^4/4) ~, \quad N=\frac{1}{\int_{-\infty}^\infty dx\, \exp( -b x^2/2 -x^4/4)} ~,
\end{equation}
where the overall normalization constant $N$ is chosen so that $a_0=1$. 
We will sometimes refer to these moments as expectation values, and alternatively denote them by $a_n \equiv \langle x^n \rangle$.
We will say that we have solved the problem associated to the measure $d\mu$ if we have computed all the moments of the distribution to some target precision. 
Instead of computing the moments by direct integration, we will review the bootstrap method of determining the $a_n$ with positivity and convex optimization. 

The first ingredient is an exact relation between the moments. In our toy model, these arise from the vanishing of a total derivative,
\begin{equation}
\int_{-\infty}^\infty dx\, \partial_x \left(x^n \exp( -b x^2/2 -x^4/4)\right) = 0 ~,
\end{equation}
so that, after expanding the derivative inside the integral, we obtain
\begin{equation}
    n a_{n-1} - b a_{n+1}- a_{n+3} = 0 ~.
\end{equation}
Moreover, since the measure is even, we can immediately set $a_{2k+1}=0$. 
With the recursion, once we determine $a_2$, and since $a_0=1$, we can determine the rest of the moments of the distribution. 
That is, we will claim that the problem is solved if we have determined $a_2$ for some value of $b$ to some high level of precision.

The second ingredient is some form of positivity. 
In our toy model, the idea is that given any polynomial of $x$ with arbitrary real coefficients, $P(x)$, the integral of its square against our measure is positive definite,
\begin{equation}\label{eq:toy1 squarepos}
    \int_{-\infty}^\infty dx \, P(x)^2 \exp( -b x^2/2 -x^4/4)\geq 0 ~.
\end{equation}
This is often referred to as square-positivity in recent bootstrap literature. 
Expanding an arbitrary polynomial of finite degree $k$, $P(x) = \sum_{n\leq k} c_n x^n$, we find that the following Hankel matrix (quadratic form) built from the moments of $x$ is positive semi-definite (PSD)\footnote{Indeed, expanding \eqref{eq:toy1 squarepos} we obtain $N\sum_j \sum_k c_j c_k \int dx\, x^{j+k} \exp( -b x^2/2 -x^4/4) \equiv c^{\text{T}}\mathcal{M}^{(k)}c\geq 0$, where $c=(c_1,\ldots,c_k)^{\text{T}}$ and $(\mathcal{M}^{(k)})_{i,j}=a_{i+j}$ is the moment matrix. The condition $c^{\text{T}}\mathcal{M}^{(k)}c\geq 0$ is equivalent to the positive semi-definiteness of $\mathcal{M}^{(k)}$ in \eqref{eq:toy1 psd momentmat}.}
\begin{equation}\label{eq:toy1 psd momentmat}
    \mathcal{M}^{(k)}= \begin{pmatrix} 1 & a_1 & a_2&\dots & a_k\\
    a_1 & a_2 & a_3 & \dots &a_{k+1}\\
    a_2 & a_3 & a_4 & \ddots &\vdots\\
    \vdots & \ddots & \ddots &\ddots& \vdots\\
    a_k & a_{k+1} & \dots & \dots & a_{2k}
    \end{pmatrix} \succeq 0 ~.
\end{equation}
Positivity implies that all the eigenvalues of $\mathcal{M}^{(k)}$ for any $k$ are positive. Having a positive $\mathcal{M}^{(k)}$ for all $k$ is not only necessary, but also a sufficient condition to recover the measure $d\mu$ from the moments (see \cite[theorem 3.8]{schmudgen2017moment} -- The solution of the Hamburger problem ).

If we now also take into account that the odd moments $a_{2s+1}$ vanish and that all the even moments $a_{2s}$ are linear functions of $a_2$, we find that 
\begin{equation}
    \mathcal{M}^{(k)} = A+ B a_2 ~.
\end{equation}
where $A,B$ are fixed matrices (that depend on the fixed value of the parameter $b$). In particular, $\mathcal{M}^{(k)}$ is linear in $a_2$ and hence is suitable for a simple convex optimization of the possible values of $a_2\in\mathbb{R}$.
Notice also that we generally expect that asymptotically the moments $a_n$ will grow. This can be estimated with a saddle point evaluation of the integral as follows
\begin{equation}
     \int_{-\infty}^\infty dx  x^{2n}\exp( -b x^2/2 -x^4/4)=\int_{-\infty}^\infty\exp(-bx^2/2 -x^4/4 +n \log(x^2))
\end{equation}
The saddle (for the variable $u=x^2$) occurs when 
\begin{eqnarray}
\partial_u( -b u/2- u^2/4+n\log(u))=0\\
-b/2- u/2+n/u=0
\end{eqnarray}
 When $n$ is very large $ u \simeq \sqrt{2n}$   and we see that the dominant term comes from the logarithm which  grows like  $\frac 12n \log(n) $. This indicates that the sequence $a_{2n}$ eventually grows similarly to $n!$.

Specifically, we can now consider the following two semi-definite optimization problems (SDP programs):
\begin{eqnarray}\label{eq:toy1 SDP}
\text{SDP:}\quad
a_{2}^{(k),+} &=& \text{max}(a_2 | \mathcal{M}^{(k)} \succeq 0)\\
a_{2}^{(k),-} &=& \text{min}(a_2 | \mathcal{M}^{(k)} \succeq 0)
\end{eqnarray}
These two provide a rigorous upper ($+$) and lower ($-$) bound on the value of $a_2$.
It is easy to prove that since positivity of $\mathcal{M}^{(k)}$ is a subset of conditions on the positivity of $\mathcal{M}^{(k+1)}$, that $a_{2,+}(k)$ is a descending sequence. Similarly, $a_{2,-}(k)$ is an ascending sequence. 
If we truncate the problem numerically to some fixed value $k$, we will say that we have solved/optimized the constraints to level $k$.
The window of allowed values for $a_2$, $\Delta a_2^{(k)}= a_{2}^{(k),+}-a_{2}^{(k),-}$, may also be thought of as a rigorous error bar. 

We illustrate this with an example for $b=1$ in table \ref{tab:fast-conv}, where we observe exponentially fast convergence to the true value, which can be determined with other analytic methods to give $a_2(b=1)= 0.4679199170\ldots$. Indeed, consider $Z[b]=\int_{-\infty}^\infty dx \exp( -b x^2/2 -x^4/4)$. This can be expressed in terms of the Bessel $K$ function as
\begin{equation}
Z[b]=\sqrt{\frac{b}{2}} \, \e^{\frac{b^2}{8}} K_{\frac{1}{4}}\left(\frac{b^2}{8}\right)
\end{equation}
and thus $a_2(b)= 2\partial_b \log(Z[b])$, which can be readily evaluated numerically to very high precision.

\begin{table}[h!]
\centering
\begin{tabular}{|c|c|c|c|}\hline
Level $k$ & $a_{2}^{(k),-}$ & $a_{2}^{(k),+}$ &$\Delta a_2^{(k)}$\\
\hline
5  & $0.43425855$ & $0.47602360$ & $0.00236958$ \\
7  & $0.46600537$ & $0.46837495$ & $0.00056337$ \\
9  & $0.46781157$ & $0.46794580$ & $0.00013423$ \\
11 & $0.46791372$ & $0.46792141$ & $7.692 \times 10^{-6} $\\
13 & $0.46791956$ & $0.46792000$ & $2.166 \times 10^{-7} $\\
\hline
\end{tabular}
\caption{Upper and lower bounds $a_{2}^{(k),\pm}$, and the allowed window $\Delta a_2^{(k)}= a_{2}^{(k),+}-a_{2}^{(k),-}$, for the allowed values of the moment $a_2$ for a fixed value of the parameter $b=1$ and for increasing level of truncation $k$ (the size of the moment matrix).}\label{tab:fast-conv}
\end{table}

The natural question for us is why does the result converge so quickly? This has been observed in other similar problems (see for example \cite{Berenstein:2022unr} and references therein).\footnote{In several recent bootstrap examples \cite{Kazakov:2021lel,Cho:2023ulr}, it has been shown that the SDP-based bootstrap procedure in fact converges. In this paper, however, we focus on understanding why it converges so rapidly in particular examples.} 
Our goal is to understand what mathematical property of the problem under study guarantees the fast convergence.

\paragraph{Toy example on the circle.}
We can also consider a similar trigonometric problem  on a circle, with the following periodic probability distribution
\begin{equation}\label{eq:toy1 angular model}
d\mu \propto  d\theta \exp( \beta \cos(\theta)) ~,
\end{equation}
where $\theta$ takes values in the range $-\pi\leq \theta<\pi$ and $\beta$ is a real parameter. Here, it is more natural to consider the Fourier coefficients of the distribution
\begin{equation}
    a_n = \mathcal{N} \int_{-\pi}^\pi d\theta \, \exp(i n \theta)  \exp( \beta \cos(\theta)) ~, \quad\quad  \mathcal{N}=\frac{1}{\int_{-\pi}^\pi d\theta \, \exp( \beta \cos(\theta))} ~,
\end{equation}
again normalized so that $a_0=1$. 
These moments, or Fourier coefficients, serve as a toy model for the expectation values of Wilson loop operators (or plaquettes) in more realistic physical contexts. 

Just as before, we can obtain a linear recursion for these moments by using a total-derivative relation of the form
\begin{equation}
    N \int_{-\pi}^\pi d\theta \, \partial_\theta \left(\exp(i n \theta)  \exp( \beta \cos(\theta))\right) = 0 ~,
\end{equation}
which now translate to
\begin{equation}\label{eq:toy2 totalderiv rel}
n a_n + (\beta/2) a_{n+1} -(\beta/2) a_{n-1}=0 ~,
\end{equation}
or equivalently $a_{n+1}= a_{n-1}- 2 \beta^{-1} n a_n$. As a consequence of reflection symmetry under which $\theta\to -\theta$, the $a_n$ are real-valued and $a_{-n}=a_n$. 

The positivity condition is slightly different in this case. 
The moment matrix that must be positive semi-definite is of Toeplitz form,\footnote{An elementary proof starting from square-positivity is as follows.
Instead of considering the square of an exponential operator $\e^{in\theta}$, consider the expectation value of the square of the ``cosine" operator, which is indeed non-negative
$
\int_{-\pi}^{\pi}d\theta \, \left( \cos m\theta \right)^2 \e^{\beta\cos\theta} \geq 0.$
More generally, an arbitrary linear combination of such cosine operators, $\mathcal{O} = \sum_{j} c_j \cos j\theta$, has a non-negative expectation value and thus
\begin{align}
\langle  \mathcal{O}^2 \rangle = \sum_j \sum_k c_j c_k \int d\theta\, \cos j\theta \cos k\theta \, \e^{\beta\cos\theta} 
= \sum_j \sum_k c_j c_k \frac{1}{2}\left( a_{j+k} + a_{j-k} \right) 
\equiv c^{\mathrm{T}} \mathcal{M}^{(1)} c \geq 0 ~, 
\end{align}
and hence the moment-matrix 
$
\mathcal{M}^{(1)}_{j,k} \equiv \frac{1}{2}\left( a_{j+k} + a_{j-k} \right) \succeq 0 
$
is PSD.
Similarly, by considering the expectation value of the square of an arbitrary linear combination of ``sine" operators, we obtain that a second moment matrix
$\mathcal{M}^{(2)}_{j,k} = \frac{1}{2}\left( -a_{j+k} + a_{j-k} \right) \succeq 0$.
Finally, the addition of these two, 
$\mathcal{M}_{j,k} \equiv (\mathcal{M}^{(1)} + \mathcal{M}^{(2)})_{j,k} = a_{j-k} \succeq 0$, is also PSD and equal to \eqref{eq:fourier_mat}.
}
\begin{equation}
 \mathcal{M} = \begin{pmatrix}
    1 & a_1 & a_2 &\dots\\
    a_{1} & 1& a_1 & \ddots\\
    a_{2} & a_{1} & 1 & \ddots\\
    \vdots & \ddots &\ddots &\ddots
 \end{pmatrix}\succeq 0\label{eq:fourier_mat}
\end{equation}
and require that it be positive (at some truncation level $k$). 
Using the total-derivative recursion \eqref{eq:toy2 totalderiv rel}, once $a_1$ is determined, the rest of the moments (or Fourier modes) follow. 
Clearly, if the Fourier modes of the distribution are known, the distribution itself can be reconstructed.

Two-sided rigorous bounds can be computed by solving the following SDP optimization problem for a given truncation level $k$, with a PSD matrix $\mathcal{M}^{(k)}$ in \eqref{eq:fourier_mat} of size $k$,
\begin{eqnarray}\label{eq:toy2 SDP}
\text{SDP:}\quad
a_{1}^{(k),+} &=& \text{max}(a_1 | \mathcal{M}^{(k)} \succeq 0)\\
a_{1}^{(k),-} &=& \text{min}(a_1 | \mathcal{M}^{(k)} \succeq 0) ~.
\end{eqnarray}
Notice that for this toy model on the circle, we have the more elementary bound on the moments $|a_n|\leq 1$ for all $n$. Based on this simpler bound, we will discuss a simpler bootstrap approach in the next section. 

\begin{table}[h!]
\centering
\begin{tabular}{|c|c|c|c|}\hline
Level $k$ & $a_{1}^{(k),-}$ & $a_{1}^{(k),+}$ &$\Delta a_1^{(k)}$\\
\hline
3  & $0$          & $0.78077641$ & $0.78077641$ \\
5  & $0.66666667$ & $0.71242489$ & $0.045758219$ \\
7  & $0.69648472$ & $0.69897661$ & $0.00249188$ \\
9  & $0.69774230$ & $0.69780694$ & $0.00006464$ \\
11 & $0.69777419$ & $0.69777512$ & $9.3551 \times 10^{-7} $\\
13 & $0.69777465796391$ & $0.69777465796393$ & $1.6764 \times 10^{-14} $\\
\hline
\end{tabular}
\caption{Upper and lower bounds $a_{1}^{(k),\pm}$, and the allowed window $\Delta a_1^{(k)}= a_{1}^{(k),+}-a_{1}^{(k),-}$, for the allowed values of the moment $a_1$, or first Fourier mode, for a fixed value of the parameter $\beta=2$ and for increasing level of truncation $k$ (the size of the moment matrix).}\label{tab:fast-conv 2}
\end{table}

\begin{figure}[h!]
    \centering
    \includegraphics[width=0.8\linewidth]{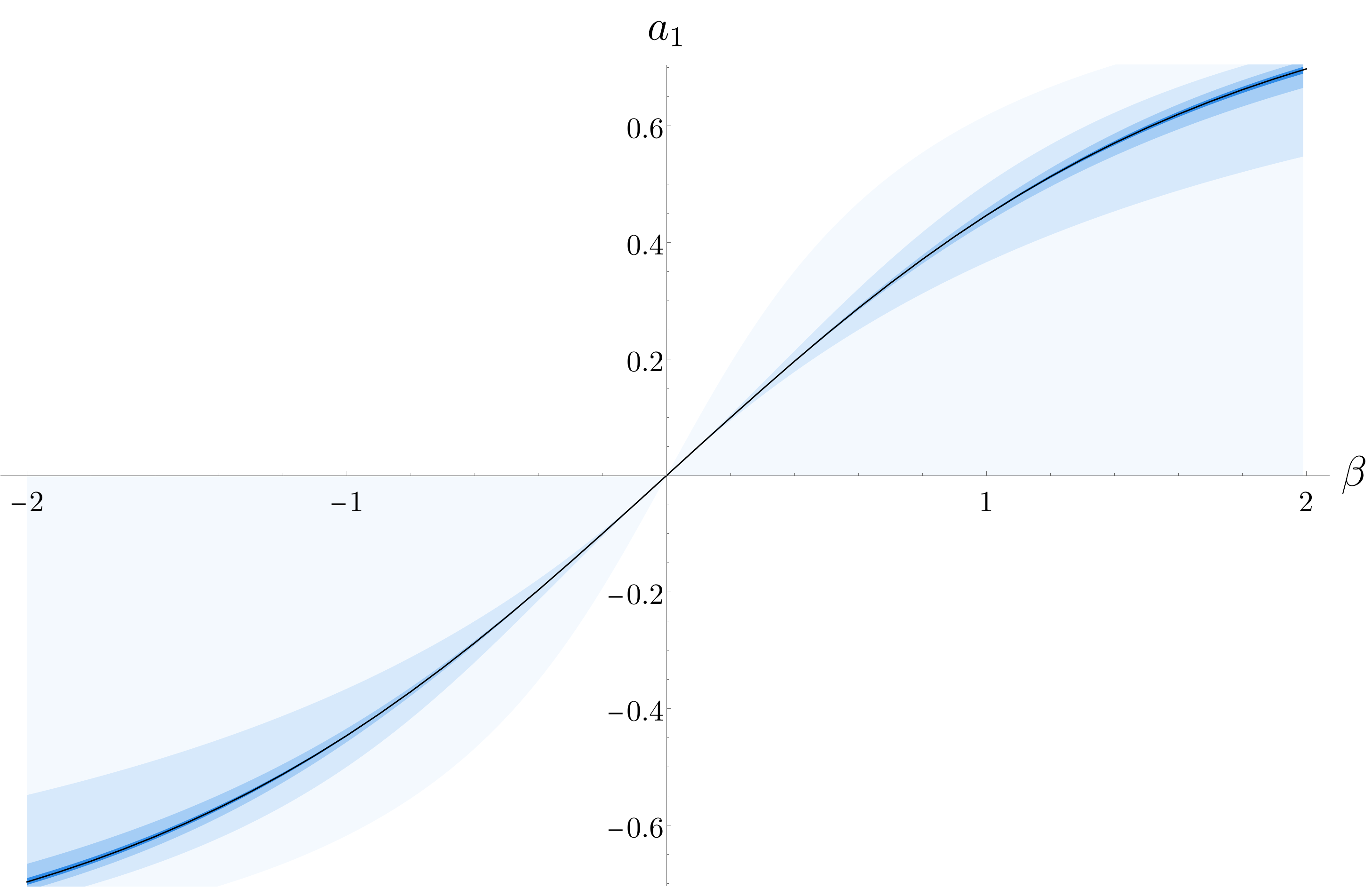}
    \caption{Results for the direct bootstrap bounds of the moment $a_1$ for a range of values of the parameter $\beta\in[-2,2]$. Shaded with increasingly opaque blue are the regions between the upper and lower bounds found by solving the SDP \eqref{eq:toy2 SDP}, for increasing truncation level $k=3,4,5,6$. The exact answer is shown in the black curve.}
    \label{fig:Toycircle1_varybeta}
\end{figure}

Table \ref{tab:fast-conv 2} shows the results obtained by solving the SDP \eqref{eq:toy2 SDP} for $\beta=2$. 
We again observe a very fast convergence to the exact value of $a_1$, which can be written in terms of the Bessel $I$ function. Defining $Z[\beta]= 2 \pi  I_0(\beta)$, we have $a_1=\partial_\beta \log(Z[\beta])|_{\beta=2}\simeq 0.697774657\ldots$.
Figure \ref{fig:Toycircle1_varybeta} shows the bootstrap bounds obtained from solving the SDP \eqref{eq:toy2 SDP} for a range of $\beta\in[-2,2]$ and increasing truncation level $k$.

These two problems are emblematic of the bootstrap program for statistical and quantum mechanical models. Our goal is to understand the mathematical properties of the bootstrap examples to improve the method. A big part of this exploration deals with the positivity of the matrices $\mathcal{M}$ and what is the best way to characterize the specific types of matrices $\mathcal{M}$ that appear in the problem. This is what we will pursue next.

\section{Problems on the real line}\label{sec:real_line}

Let us now analyze bootstrap problems on the real line. 
In particular, such problems arise in at least two broad categories. 
The first type, analogous to the toy example studied in section \ref{sec:examples}, involves integrals with respect to a measure of the form
\begin{equation}
    d\mu = \exp(- V(x))\label{eq:poly_prob}
\end{equation}
where $V(x)$ is a polynomial in $x$ such that $V(x)\to \infty$ as $x\to \pm \infty$; for concreteness, we may take $V(x) = \sum_{\ell\leq 2s} v_\ell x^\ell$ with $v_{2s}>0$. 
Similar problems can also be considered on the semi-infinite interval $(0,\infty)$. 
The second type of problem arises in one-dimensional quantum mechanics where we seek to solve the Schrodinger equation for a particle in a potential with $H=-\partial_x^2+V(x)=E$, where $E<V(\pm \infty)$ so that we are describing a proper bound state \cite{Han:2020bkb,Berenstein:2021dyf,Bhattacharya:2021btd}. We will not discuss this setting much in this paper, but the techniques are similar in that one uses positivity and finds a set of recursion relations that one needs to solve. The one difference is that one also needs to search in the value of the energy, which is actually the final target of the quantum mechanical program. In this paper we are focusing on the mathematical physics associated to the positivity of the matrices and the recursion. 

\paragraph{Total-derivative (recursion) relations and naive growth of moments.}
The idea is that in both of these types of problems there are exact, non-perturbative, recursion relations that determine higher moments of the distribution in terms of lower moments (the exact form of the recursion in the general quantum mechanical setup can be found in \cite{Berenstein:2022unr}). 
For the measure \eqref{eq:poly_prob}, these exact relations follow directly from the vanishing of a total derivative, $\int dx \, \partial_x (x^n \exp(-V(x)))=0$, which holds under the assumption that $V(x)\to \infty$ as $x\to \pm \infty$ ensuring that boundary contributions at infinity vanish. 
Expanding the derivative inside the integral, we obtain the relation\footnote{Note that the recursion \eqref{eq:poly_prob} is trivial for a Gaussian distribution as the moments are uniquely determined by the recursion and $a_0=1$, so there is nothing to bootstrap. The problem we studied in section \ref{sec:examples} is minimal in that it has the minimal number of unknowns.}
\begin{equation}\label{eq:totderiv rec}
    n\langle x^{n-1}\rangle - \langle x^n V'(x) \rangle = 0 ~.
\end{equation}
More broadly, this type of relations are often referred to as Schwinger-Dyson equations, total-derivative relations, or loop equations, depending on the context of the bootstrap problem. 
In the case of the semi-infinite interval, we would have additional boundary contributions that arise from boundary terms in integration by parts (these can be thought of as an anomaly, similar to \cite{Berenstein:2022ygg}).

What is important for us is that the coefficient of the term $n \langle x^{n-1}\rangle$ is increasing with $n$, while the rest of the coefficients in $\langle x^n V'(x)\rangle$ are of fixed size. 
The leading term in $V(x)$ is positive and of even degree $2s$, with leading coefficient $v_{2s}$,  
so we expect that $ v_{2s} \vev{x^{2s +n-1}}\simeq n \vev{x^{n-1}} $ and since $n$ is increasing we expect that the moments grow once $n\gg v_{2s}$. 
Thus, the naive expectation is that the moments grow faster than exponentially. The same is true for the quantum mechanical bootstrap problems: there is a recursion where the higher-order moments are expected to grow faster than exponentially.

\paragraph{Decay of the minimal eigenvalue of $\mathcal{M}$ and fine-tuning.}
Consider the positivity of the moment matrix at some truncation level (or size) $k$,
\begin{equation}
    \mathcal{M}^{(k)}= \begin{pmatrix} 1 & a_1 & a_2&\dots & a_k\\
    a_1 & a_2 & a_3 & \dots &a_{k+1}\\
    a_2 & a_3 & a_4 & \ddots &\vdots\\
    \vdots & \ddots & \ddots &\ddots& \vdots\\
    a_k & a_{k+1} & \dots & \dots & a_{2k}
    \end{pmatrix} \succeq 0 ~.
\end{equation}
That is, the eigenvalues of $\mathcal{M}^{(k)}$ are all non-negative for every $k$. 
Furthermore, by the lemmas in appendix \ref{sec:review of matrices and convexity}, or by the eigenvalue interlacing theorem, the following inequalities are true
\begin{equation}
    \lambda_{\text{min}}(\mathcal{M}^{(k)}) \geq \lambda_{\text{min}}(\mathcal{M}^{(k+1)}) ~.
\end{equation}
That is, the minimal eigenvalue of $\mathcal{M}^{(k)}$ \emph{decreases} as we increase $k$, whereas the entries of the matrix themselves, i.e. the moments, \emph{grow} factorially.

A more precise statement that is known from the mathematical literature is that $\lambda_{\text{min}}(\mathcal{M}^{(k)})$ decreases faster than any power law for measures on the real line, in situations where the moments uniquely determine the distribution (see for example \cite{chen1999small,berg2011smallest}). 
Essentially, the smallest eigenvalue of $\mathcal{M}^{(k)}$ becomes exponentially suppressed. One could say that the minimal eigenvalue is finely tuned, considering that the matrix elements of $\mathcal{M}$ are actually growing. 
In the context of the quantum mechanical bootstrap, if one is not at the correct energy for a bound state, one expects that there will be a failure of positivity at some $k$ and then the smallest eigenvalue is negative. This smallest eigenvalue will keep on decreasing  further as we increase $k$: the negative  eigenvalue will grow in absolute value and become a large negative number eventually.

Now let us give a quick proof of the general decay. Consider a measure $d\mu$ of the form \eqref{eq:poly_prob} that is bounded above and decays at least as fast as an exponential. In the paper \cite{chen1999small}, the authors studied the problem of the measure $d\mu_\beta \simeq \exp(-|x|^\beta) dx$ and found the decreasing asymptotic form of the eigenvalues. Consider now that due to the smoothness of $d\mu$, we can bound $d\mu$ from above by an element of $d\mu_\beta$ for some $\beta$.  That is, we can write $\kappa\,  d\mu_{\beta} \succeq d\mu $ for some $\kappa$. We can easily establish that 
\begin{equation}
    \kappa \mathcal{M}^{(k)}(\beta) \succeq \mathcal{M}^{(k)}
\end{equation}
so that the minimal eigenvalue of $\mathcal{M}^{(k)}$ must be smaller than $\kappa$ times the minimal eigenvalue of $\mathcal{M}^{(k)}(\beta)$ that is exponentially suppressed. This is not the most general proof of the smallness of the eigenvalue, as the measure $d\mu$ might have singularities that one needs to take care of (e.g. a delta function at some value), but it is valid in all the cases we are interested in. The upshot is that the smallest eigenvalue of $\mathcal{M}^{(k)}$ is asymptotically very small.

To illustrate this decay of the minimal eigenvalue, it is instructive to consider a simple example where $d\mu = \exp(-x) dx$ on $(0,\infty)$, or equivalently $d\mu = \exp(-|x|) dx$ on $\mathbb{R}$ albeit not smooth at the origin. The moments are given by $\mu_n = \Gamma(n+1)$. This is a Stieltjes problem, where there are two Hankel matrices to consider. These are
\begin{equation}
   \mathcal{M}_0 =\begin{pmatrix} a_0 & a_1 &\dots\\
    a_1&a_2&\dots\\
    \vdots& \vdots&\ddots \end{pmatrix}, \quad \mathcal{M}_1=\begin{pmatrix} a_1 & a_2 &\dots\\
    a_2&a_3&\dots\\
    \vdots& \vdots&\ddots \end{pmatrix}
\end{equation}

We can ask how does the last $2\times2$ block of $\mathcal{M}_{0,1}^{(k)}$ look like. We obtain that
\begin{equation}
    \begin{pmatrix} \mu_{n-1} & \mu_{n}\\
    \mu_{n} & \mu_{n+1}
    \end{pmatrix} 
    = \Gamma(n) \begin{pmatrix} 1& n\\
    n& n(n+1)
    \end{pmatrix}
\end{equation}
so the  largest eigenvalue of this two-by-two block grows like $\Gamma(n)\,n\,(n+1)$, while from the determinant, the smallest eigenvalue is of order $\Gamma(n) / n $, that is, it is suppressed by order $1/n^3$ relative to the largest eigenvalue. One can do further numerical experiments to find that this suppression of order $1/n^3$ occurs from eigenvalue to eigenvalue when we consider bigger matrices on the tail. Even this suppression near the tail for the large eigenvalues is useful for our arguments. It shows that the matrix is finely tuned on the submatrices near the tail of the problem.

When we vary the unknown elements of the recursion, we expect that the changes in the matrix become large. The change in the smallest eigenvalue generically should be of order the change in the matrix elements if one uses a naive scaling analysis (a physicists' order of magnitude estimate). Since we have been arguing that this small eigenvalue is a very small number already, it is extremely sensitive to changes in the matrix elements of $\mathcal{M}$. Essentially, we should expect that the error bars on the unknown elements of the recursion become exponentially small. 

We claim that this is the justification for the high accuracy of the bootstrap method. First,  the matrix elements of the positive matrix are growing factorially and in the opposite direction, the smallest eigenvalue is decreasing exponentially fast.

\paragraph{An eigenvalue bootstrap.}
This expected acute sensitivity of the minimal eigenvalue of $\mathcal{M}^{(k)}$ to the initial conditions, i.e. the possibly allowed values for the low-order moments, motivates us to consider a different algorithm\footnote{See \cite{Berenstein:2022unr} for a similar algorithm considered in the quantum mechanical bootstrap.}: Maximize the $\lambda_{\text{min}}(\mathcal{M}^{(k)})[a_m]$ as we scan in the space of the unknown moments $a_m$ (those undetermined from the total-derivative relations)\footnote{Note that this is not a very cheap algorithm, as we still have to find this minimal eigenvalue as we scan over the space of undetermined moments, but it can be implemented as an SDP problem.}. A good estimate for the true value of the undetermined moments, $\bar{a}_m$, is then that which extremizes $\lambda_{\text{min}}(\mathcal{M}^{(k)})[a_m]$.

To approximate the error in our estimate, we can examine the Hessian in the space of the undetermined moments. 
In this setup, at optimality $\lambda_{\text{min}}^*$ is at a maximum and the minimal eigenvalue is a convex function. 
We expect the Hessian $h_{ij}=\partial_{i}\partial_{j} \lambda^*_{\text{min}} $ to be negative definite and naively it is controlled by perturbation theory. The Hessian matrix should be large (controlled by the large entries in the matrix), so the quadratic approximation $\lambda_{\text{min}} \simeq \lambda_{\text{min}}^* -\frac 12 h_{ij} \delta a_i \delta a_j$ will be an approximation for $\lambda_{\text{min}}(a)$ as a function of the unknowns. 
This gives an ellipsoid bound of the size of the error bars.

This is, however, not the complete story. Let us return to the problem in section \ref{sec:examples}, where
$$
d\mu\propto \exp(-x^2/2-x^4/4) 
$$
The matrix of moments is such that $a_{2k+1}=0$ because it is an even function; that is, the measure has a $\mathbb{Z}_2$ symmetry under which $x\to-x$. 
In fact, the positivity of $\mathcal{M}^{(k)}$ is equivalent to the positivity of two blocks, with matrices given by\footnote{This is one of the simplest examples of \emph{invariant SDP} \cite{bachoc2012invariant}, which in this case implies that it is sufficient to impose the positive-semi-definiteness of two moment matrices constructed from ``states" transforming in the corresponding irreducible representation of the symmetry group $\mathbb{Z}_2$ (even and odd). In our toy model, these two moment matrices are built from even powers of $x$ and odd powers of $x$, respectively: $(\mathcal{M}_\text{e})_{i,j}=\langle x^{i+j}\rangle$ with $i,j\in2\mathbb{Z}_{\geq0}$ and $(\mathcal{M}_\text{o})_{i,j}=\langle x^{i+j}\rangle$ with $i,j\in2\mathbb{Z}_{\geq0}+1$}
\begin{equation}
    \mathcal{M}_{\text{e}}= \begin{pmatrix}
        1& a_2 &a_4&\cdots\\
        a_2 & a_4& a_6 &\ddots\\
        a_4 & a_6 &a_8 &\ddots\\
        \vdots & \ddots & \ddots &\ddots
    \end{pmatrix} ~, 
\quad \mathcal{M}_{\text{o}} =\begin{pmatrix}
         a_2 &a_4&\cdots\\
         a_4& a_6 &\ddots\\
        \vdots  & \ddots &\ddots
    \end{pmatrix} ~,
\end{equation}
where the subscript ``e" (o) denotes that the moment matrix is built from monomials in $x$ with even (odd) powers. 
In particular, the minimal eigenvalue of $\mathcal{M}^{(k)}$ is the minimal eigenvalue between the two matrices $\mathcal{M}^{(k)}_{\text{e}}$ and $\mathcal{M}^{(k)}_{\text{o}}$.

\begin{figure}[h!]
    \centering
    \includegraphics[width=0.475\linewidth]{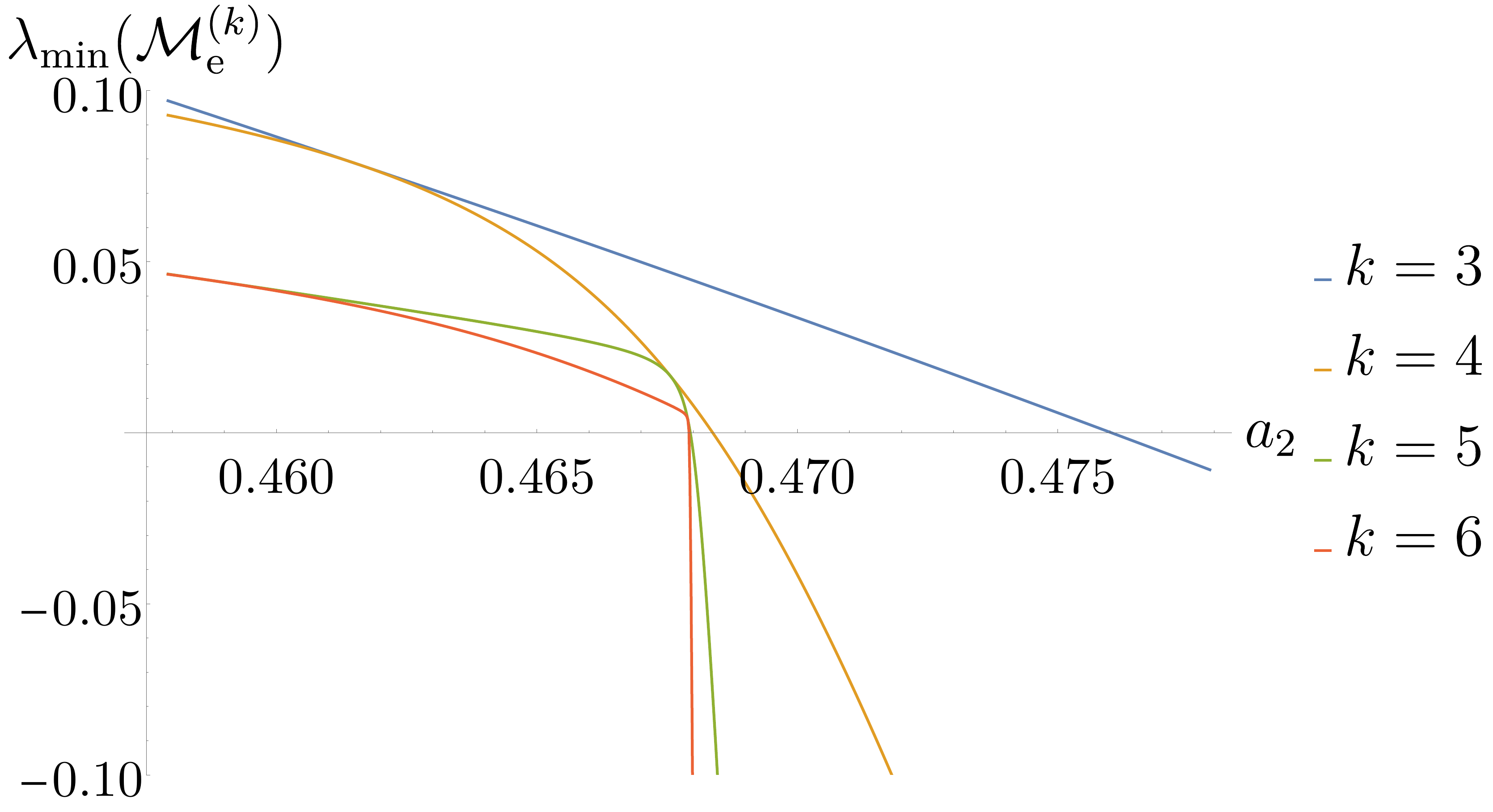} ~
    \includegraphics[width=0.475\linewidth]{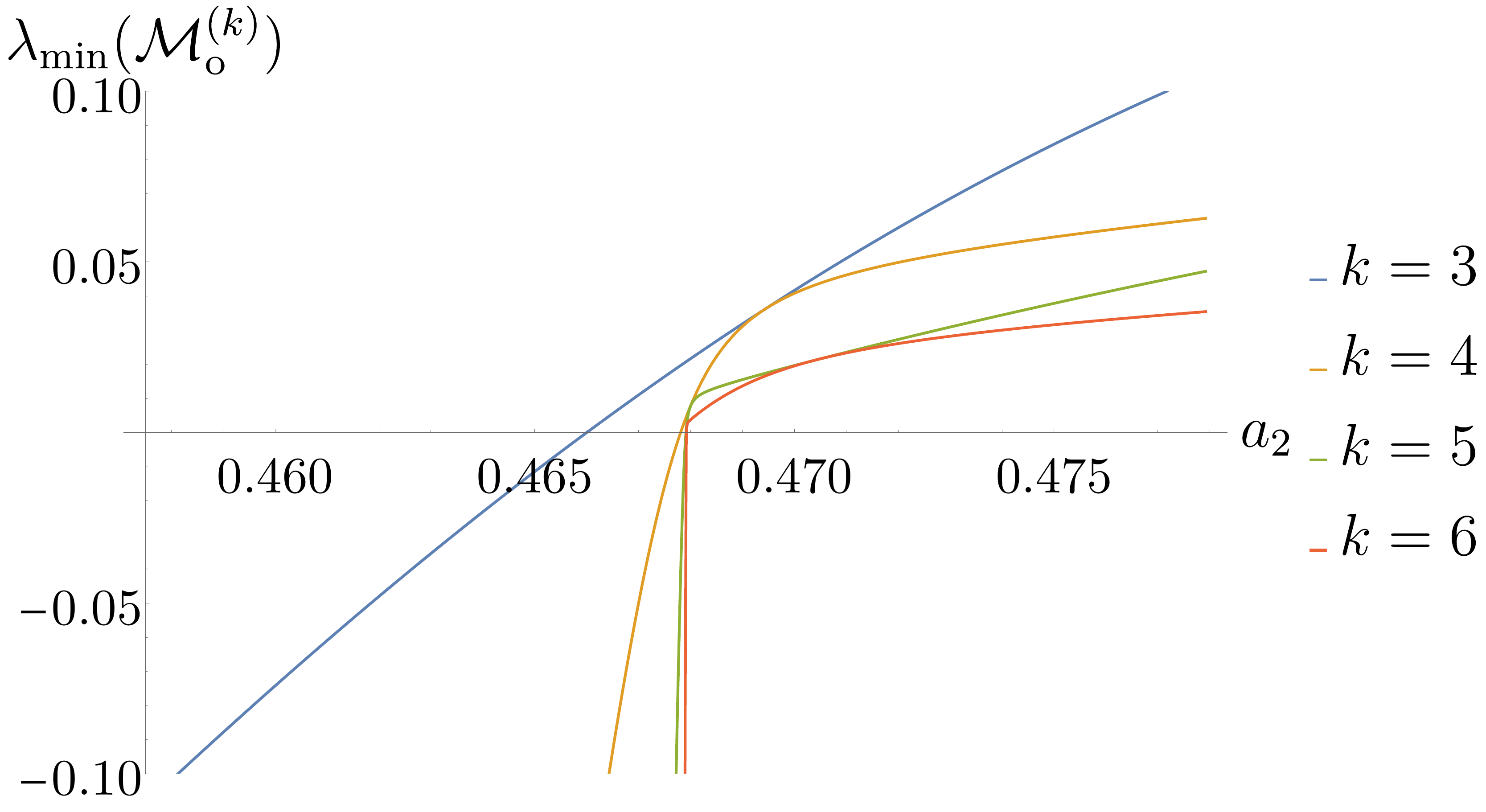}
    \caption{Minimum eigenvalue of the PSD matrices $\mathcal{M}^{(k)}_{\text{e}}$ (left) and $\mathcal{M}^{(k)}_{\text{o}}$ (right) as a function of the unknown moment $a_2=\langle x^2 \rangle$, for increasing truncation size $k=3,4,5,6$ of the moment matrices.}
    \label{fig:toyog mineval each M}
\end{figure}

\begin{figure}[h!]
    \centering
    \includegraphics[width=0.8\linewidth]{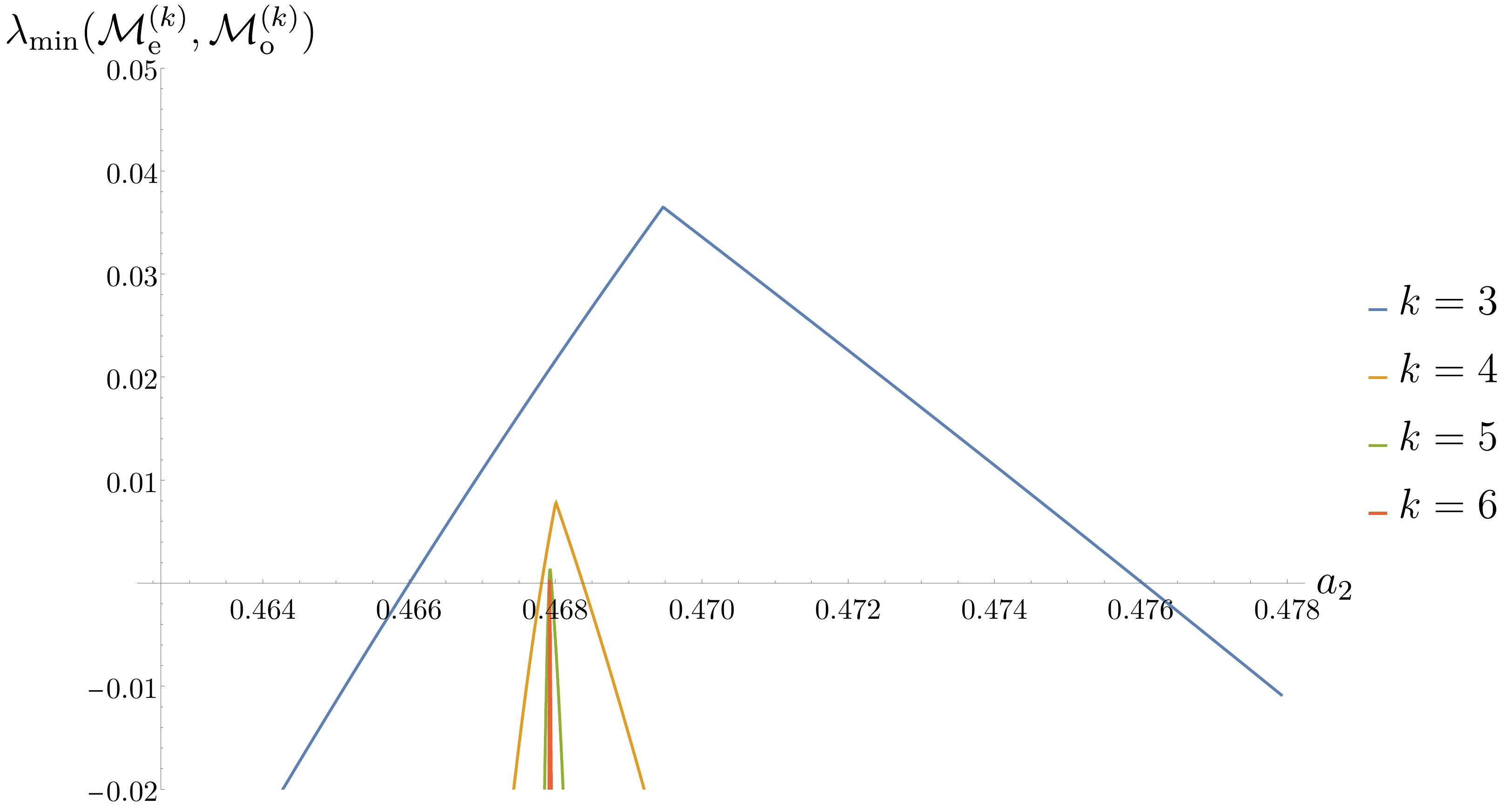}
    \caption{Minimum eigenvalue $\lambda_{\text{min}}(\mathcal{M}^{(k)}_{\text{e}},\mathcal{M}^{(k)}_{\text{o}})$ of both PSD matrices $\mathcal{M}^{(k)}_{\text{e}}$ and $\mathcal{M}^{(k)}_{\text{o}}$ as a function of the unknown moment $a_2=\langle x^2 \rangle$, for increasing truncation size $k=4,5,6,7$ of the moment matrices.}
    \label{fig:toyog mineval}
\end{figure}    

As can be seen from figure \ref{fig:toyog mineval}, the maximum of the minimum eigenvalue is not at a smooth point of the curve. What is going on is that the eigenvalue associated to the maximum is degenerate between the two matrices and the Hessian is not well defined (see figure \ref{fig:toyog mineval each M}). Instead, there are first derivatives to the left and to the right of the maximum that can be used to define exact upper bounds for the ranges. Because of convexity, the curve of the minimum eigenvalue must be exactly below the straight lines emanating from the maximum that are tangent to the minimal eigenvalue curve.
This is a special case for when the distribution is even. The existence of the symmetry allows for level crossings on the matrix $\mathcal{M}^{(k)}$ as a function of parameters. It is curious that the maximum is exactly at one of these points.

The point we are making is that the Hessian matrix of perturbation theory is not necessarily a good guide for how the bounds on the undetermined moments will behave. 
Instead, to approximate the error in our estimate from our eigenvalue  algorithm we can simply take the difference between the estimate at level $k$ and that at level $k-1$. In our toy model, shown in figure \ref{fig:toyog mineval}, we indeed observe that this is indeed a more reasonable approximate error bar. 

Finally, let us contrast this eigenvalue bootstrap estimate and its approximate error bar with the following simple rigorous bound. 
The positivity of $\mathcal{M}^{(k)}\succeq0$ in particular implies that $\lambda_{\text{min}}(\mathcal{M}^{(k)})\geq 0$, and thus the equality for the latter provides a sharp bound on the allowed region in the space of undetermined moments. 
We can also observe this for our toy model in figure \ref{fig:toyog mineval}, and furthermore notice that the interval where $\lambda_{\text{min}}(\mathcal{M}^{(k)})\geq 0$ shrinks rapidly with increasing $k$. Again, this latter fact we understand as a consequence of the acute sensitivity of the eigenvalues of $\mathcal{M}^{(k)}$ to the possible values for the undetermined moments $a_m$, discussed in the previous paragraph. 
While this simple sharp bound might be easy to calculate in bootstrap problems with one free variable, in more complicated problems with many undetermined moments (variables) our estimate and its approximate error bar from the drift in its center value might prove more useful. The algorithm finds one interior point and we expect that the regions of allowed parameters shrink exponentially fast, so we do not have to find separate bounds on all the variables.

\paragraph{A bootstrap problem that converges slowly.}
We end this section by noting an example of a bootstrap problem similar to those studied above that actually converges slowly. This means that the problem only converges polynomially fast, rather than exponentially fast. 
Let us setup a problem where we work on the interval $(0,\infty)$ and we use a measure $d\mu =b \delta(x) +\exp(-x)$. We have added a delta function at the origin that can only modify the zero-th moment of the distribution.
The measure is positive only if $b>0$, but let us now set it up in a bootstrap sense. We find that
$a_0=1+b$, and $a_s= \Gamma(s+1)$. We need to check positivity of the matrix
\begin{equation}
    \mathcal{M} =  \begin{pmatrix} b+1 & 2& 6 &\cdots\\
    2 & 6 & 24 &\ddots\\
    6& 24& $5!$ &\ddots\\
    \vdots & \ddots & \ddots &\ddots
    \end{pmatrix}
\end{equation}
at various levels. Since we know that the bottom corner starting at matrix element $\mathcal{M}_{2,2}$ is already positive, we just need to check how adding $b$ changes the problem. A simple way to do this is to notice that $\det(M_k)$ is linear in $b$ and must be positive. It can only change sign when the smallest eigenvalue goes negative. The solution for $b$ at a given level is solved by setting 
$\det(\mathcal{M}^{(k)}(b))=0$. We find by direct numerical computation that $b(k)\geq -1/k$ for matrices of size $k\times k$. That is,
we find that $b_{\text{min}}\to 0 $ as $k\to \infty$ only as $1/k$. This runs counter to our statements about 
generic exponential convergence. The issue here is that if we use perturbation theory on the smallest eigenvalue, evaluated at $b=0$ for some fixed $k$, the eigenvector associated to this small eigenvalue has a suppressed amplitude in the first entry. This means that the perturbation of the small eigenvalue due to $b$ is also suppressed and competes with the smallness of the eigenvalue itself. This case is obviously somewhat extreme, as we concentrated the bootstrap problem in one moment. Again, our purpose here is to say that although one generically expects exponentially fast convergence for bootstrap problems of polynomial type, this needs to be analyzed in detail for any problem of interest. 

\section{Problems on the circle }\label{sec:circle}

Let us turn to trigonometric integrals with a measure $d\mu(\theta)$ of the form
\begin{align}\label{eq:model on the circle}
\int_{-\pi}^{\pi} d\mu(\theta) = \int_{-\pi}^{\pi} d \theta \, \exp({-V(\theta)}) ~,
\end{align}
where $V(\theta)=V(\theta+2\pi)$ is a real-valued periodic potential, and where the positive semi-definite moment (or Fourier mode) matrix takes a Toeplitz form,
\begin{align}
(\mathcal{M})_{j,k} = \frac{1}{2}\left( a_{j-k} + a_{-(j-k)} \right) \succeq 0 ~,
\end{align}
where 
\begin{equation}\label{eq:def trig moments}
    a_n = N \int_{-\pi}^\pi d\theta \, \exp(i n \theta)  \exp({-V(\theta)}) ~, \quad\quad N=\frac{1}{\int_{-\pi}^\pi d\theta \, \exp({-V(\theta)})} ~.
\end{equation}
For simplicity, in the following we will assume that the potential $V(-\theta)=V(\theta)$ is even, and so the PSD matrix takes the form
\begin{equation}
\mathcal{M}= \begin{pmatrix}
    1 & a_1 & a_2 &\cdots\\
    a_{1} & 1& a_1 & \ddots\\
    a_{2} & a_{1} & 1 & \ddots\\
    \vdots & \ddots &\ddots &\ddots
 \end{pmatrix}\succeq 0 ~.
\end{equation}
Our goal is now to understand positivity in a way similar to the moment problem that we studied in the previous section. 
The idea again is to try to understand the positivity of $\mathcal{M}$ in terms of the eigenvalues of the matrix. 

\paragraph{Moments as Fourier modes.}
The first observation is that the moments $a_n$ defined in \eqref{eq:def trig moments} are, up to an overall normalization, nothing but the Fourier modes of the measure $d\mu$. 
The toy model \eqref{eq:toy1 angular model} is simple since in this case we know exactly these Fourier modes in terms of the modified Bessel function of the second kind:
\begin{align}
\e^{\beta \cos(\theta)} = \sum_{n\in\mathbb{Z}} I_n(\beta) \e^{in\theta} ~.
\end{align}
In particular, the $a_n$ decay (faster than) exponentially as $n\to\infty$.

More generally, from simple Fourier analysis, the decay of the Fourier modes as $n\to\infty$ is controlled by the smoothness of the measure $d\mu$; for smooth, or infinitely-differentiable $V(\theta)$ this decay is exponential. 
However, our conventional bootstrap procedure of solving an SDP imposing $\mathcal{M}\succeq0$ is \emph{agnostic} about this exponentially suppressed behavior of high Fourier modes, or moments.

\paragraph{Minimum eigenvalue as the minimum of the measure.}
As discussed in the previous section, a eigenvalue bootstrap procedure is to look at regions where the minimum eigenvalue of the PSD matrix $\mathcal{M}$ is positive. For measures defined on $\mathbb{R}$, this is not very practical as we still have to compute this minimum eigenvalue. However, in angular integrals we can exploit the symmetry of the moment matrix $\mathcal{M}$ to deduce this minimum eigenvalue as follows. 

First, note that by the expected strong decay of large moments $a_n$ as $n\to \infty$, the PSD moment matrix takes the approximate banded form 
\begin{align}\label{eq:approx circulant}
\mathcal{M}
\simeq \begin{pmatrix}
        1 & a_1  & a_2  & \cdots & a_{n_0}  & 0 & 0 & \cdots & 0 \\
        a_1  & 1 & a_1  & \cdots & a_{n_0-1}  & a_{n_0} & 0 & \cdots & 0 \\
        \vdots & a_1 & 1 & &  &  & \ddots & & \vdots \\
        a_{n_0} & \vdots &  & \ddots  \\
        0 & a_{n_0} &  \\
        0 & 0 & \ddots \\
        \vdots & \vdots &  & \\
        0 & 0 & \cdots  \\
    \end{pmatrix} ~,
\end{align}
where we have set the moments $a_{m}=0$ with $m>n_0$ for a sufficiently large integer $n_0$. 
Away from the upper-left and lower-right corners of \eqref{eq:approx circulant}, this matrix is approximately a circulant matrix, whose eigenvalues are given by
\begin{align}\label{eq:approx eval of circulant mat}
\lambda_j = \sum_{|m|\leq n_0} a_m \, \e^{\frac{2\pi i}{L}jm} ~,
\end{align}
where $j=1,\ldots,L$ labels the different eigenvalues of a $L\times L$ matrix. 
This is a good symmetry away from the two corners of the matrix, which is like translation invariance in a lattice system. 
The violation of the symmetry at the corners can be thought of as a boundary condition at the edges of an interval in the corresponding  $1$d lattice problem. 

\begin{figure}[h!]
    \centering
    \includegraphics[width=0.65\linewidth]{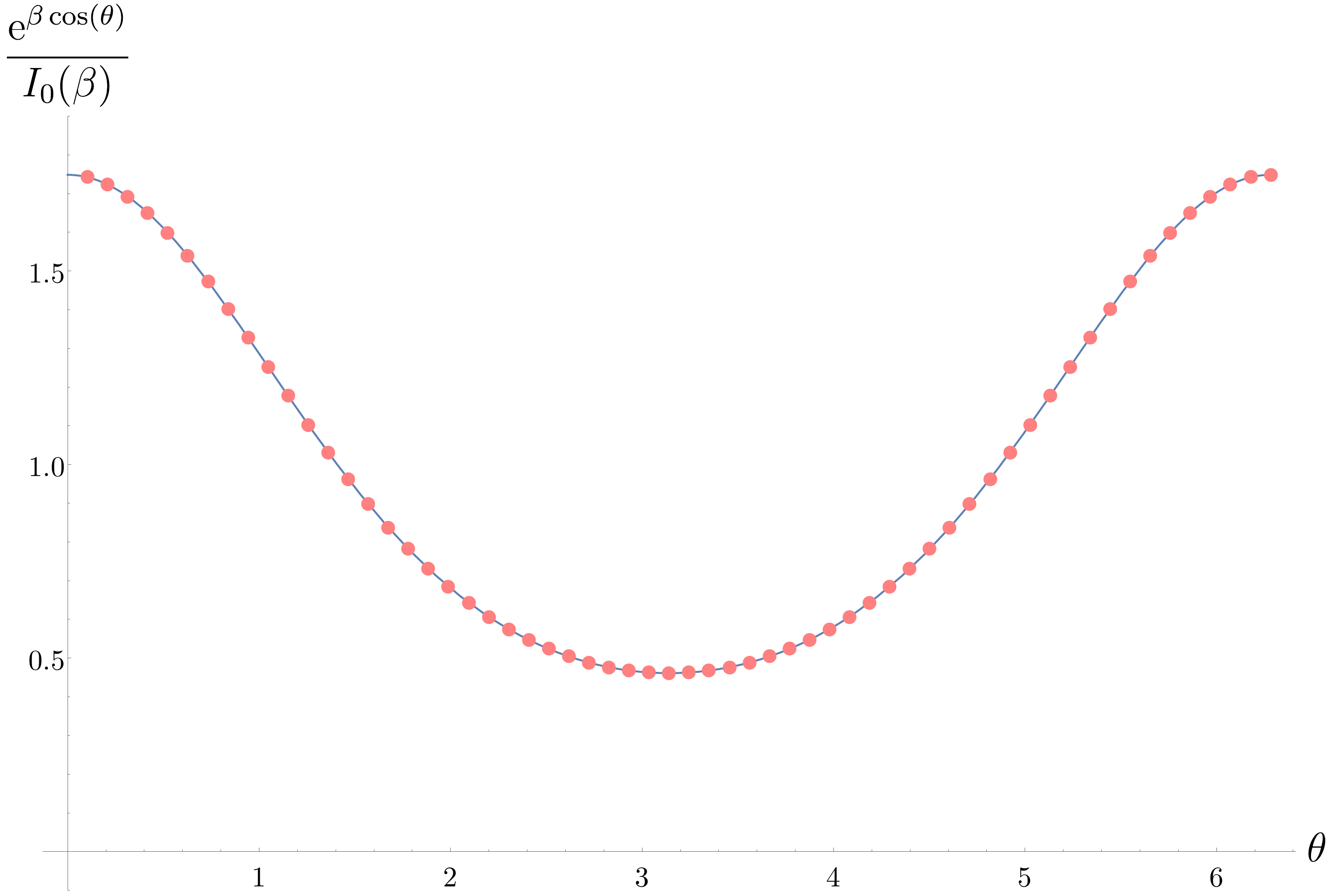}
    \caption{Example of how the eigenvalues \eqref{eq:approx eval of circulant mat} of a circulant matrix reconstruct the measure $d\mu(\theta)$. Shown in blue is the plot of the normalized measure of the toy example $d\mu(\theta) = I_0(\beta)^{-1}\,\e^{\beta\cos\theta}$ for a choice of $\beta=\frac{2}{3}$. Shown in pink are $\frac{2\pi}{L}\times\lambda_j$ where $\lambda_j$ are the eigenvalues \eqref{eq:approx eval of circulant mat}, which approximate the eigenvalues of the PSD Toeplitz matrix $\mathcal{M}$, for a choice of $L=60$. In this example, the moments $a_m$ for $|m|\leq n_0=20$ in \eqref{eq:approx eval of circulant mat} where computed using the shoestring  bootstrap method discussed in section \ref{subsec:the cheap bootstrap}.}
    \label{fig:reconstruct measure from evals example}
\end{figure}

As noted in the previous paragraph, the moments $a_n$ are Fourier modes of the measure $d\mu(\theta)$. In view of \eqref{eq:approx eval of circulant mat}, which itself is a Fourier transform, we learn that the eigenvalues \eqref{eq:approx eval of circulant mat} reconstruct the measure $d\mu$ itself. 
Figure \ref{fig:reconstruct measure from evals example} shows an example of how the eigenvalues \eqref{eq:approx eval of circulant mat}, which approximate the eigenvalues of the PSD Toeplitz matrix $\mathcal{M}$, accurately reconstruct the measure of the toy model \eqref{eq:toy1 angular model}.

In particular, the minimum eigenvalue --- whose region of positivity in moment space provides a rigorous bound for the undetermined moments $a_m$, and whose maximization in moment space we discussed previously as the eigenvalue bootstrap approach --- precisely corresponds to the minimum of the measure $d\mu(\theta)$, or maximum of the potential $V(\theta)$, itself.

In particular, for the model \eqref{eq:model on the circle} we find that the minimal eigenvalue of $\mathcal{M}$ should be similar to the minimal value of $d\mu$ on the circle and this is a (fixed) positive number. 
That is, we find that the matrix $\mathcal{M}$ has a gap on the eigenvalues relative to zero. The positivity bounds that we would derive are not close to the eigenvalues we would expect from $d\mu$. The bound on the quantities arises exactly when the PSD matrix $\mathcal{M}$ has the last eigenvalue equal to zero. In problems where we already know the minimum of the distribution, we could improve the algorithm by taking this gap into account.

\paragraph{Goldilocks.} Thus far we have only discussed properties of the moments $a_n$ and of the PSD moment matrix $\mathcal{M}$. Let us now consider the structure of our exact total derivative relations between the moments. 

Consider the total derivative relations of our toy model \eqref{eq:toy1 angular model}: $a_{n+1} = -\frac{2}{\beta} n a_n + a_{n-1}$. 
The coefficient of the first term on the RHS of this recursion relation suggests that $a_n$ grows factorially as $n\to\infty$. More precisely, when expressing $a_n$ for large $n$ in terms of the first few undetermined moments $a_1, a_2,\ldots$ (in our toy example, just $a_1$), the coefficients of the recursion relation grow factorially.
Based on this recursion relation, we expect $a_n$  itself to grow factorially unless a delicate \emph{fine-tuning} occurs among the remaining terms, which also have large coefficients in the recursion. 
Since the moments satisfy $|a_n|\leq 1$, we conclude that such fine-tuning must indeed be present.
In other words, the large-$n$ values of $a_n$ exhibit a very strong sensitivity to the “initial conditions” --- that is, the assumed values of the low-order undetermined moments $a_1, a_2,\ldots$. 

Turning this argument around explains why the bootstrap approach based on SDP, or on the positivity of the minimum eigenvalues, converges very quickly for this class of models: the factorial growth of the recursion coefficients makes it very easy to violate the condition $|a_n|<1$ for large $n$ (let alone the expected exponential decay discussed in the previous paragraph) as we scan over the possible values of the low moments.

Similar Toeplitz matrices arise in other bootstrap problems associated to spin chains, where convergence of the SDP was generally poor for data that was not the target of the SDP problem \cite{Berenstein:2024ebf}. The $a_n$ are correlation functions at finite distance $n$ in a lattice QFT. We expect that similarly to this case, the correct solution the $a_n$ are decaying rapidly and one could expect that the correct vacuum Toeplitz matrix has a gap in the eigenvalues relative to zero, where the solution aims to get to the ground state of a spin chain. Unlike the setups studied here, all the $a_n$ are independent variables unless the system is free. Perhaps integrable systems are also amenable to solving the correlation functions by bootstrap ideas.
If that is the case, the edge of positivity of the matrix is far from the true solution. This means that the constraints in $a_n$ will be generally poor, perhaps explaining the poor convergence to the correlation functions. Studying this question in more detail is beyond the scope of the present paper.

\paragraph{A  cheap bootstrap, for angular integrals.} 
In fact, this sensitivity in our toy model is so strong that simply imposing $|a_n|\leq1$ for sufficiently large $n$ --- implied\footnote{This is just the positivity of the eigenvalues $1\pm a_n$ of the principal minor $\begin{pmatrix} 1 & a_n \\ a_n & 1 \end{pmatrix}$ of $\mathcal{M}$.}, in particular, by the positivity of the full moment matrix $\mathcal{M}$ --- yields very tight rigorous bounds.
To illustrate this, consider again our toy model \eqref{eq:toy1 angular model}. 
Let us denote by $a_{1}^{(m_0),\pm}$ the upper ($+$) and lower ($-$) bounds obtained by imposing $|a_{m_0}|\leq 1$ together with the total-derivative recursion $a_{n+1} = -2\beta^{-1} n a_n + a_{n-1}$.
Table \ref{tab:a 2nd cheap boot} shows the resulting bounds for increasing values of $m_0$, and for a fixed value of the parameter $\beta=2$.\footnote{Table \ref{tab:a 2nd cheap boot} should be compared to table \ref{tab:fast-conv 2} with truncation level $k=m_0+1$, since $\mathcal{M}^{(k)}$ of size $k$ includes moments up to $a_{k-1}$.}
Due to the linearity of the total-derivative recursion relations, obtaining these bounds amounts to solving a simple linear equation instead of a full SDP program. Additionally, as long as we choose rational values of $\beta$, we can compute these simple bounds to arbitrary precision without floating point error.

\begin{table}[h!]
\centering
\begin{tabular}{|c|c|c|c|}\hline
$m_0$ & $a_{1}^{(m_0),-}$ & $a_{1}^{(m_0),+}$ &$\Delta a_1^{(m_0)}$\\
\hline
4  & $\frac{3}{5}=0.6000000000$ & $\frac{4}{5}=0.8000000000$ & $\frac{1}{5}=0.2000000000$ \\
6  & $\frac{52}{75}=0.6933333333$ & $\frac{158}{225}=0.7022222222$ & $\frac{2}{225}=0.008888888889$ \\
8  & $\frac{30}{43}=0.6976744186$ & $\frac{3481}{4988}=0.6978748998$ & $\frac{1}{4988}=0.000200481155$ \\
10  & $\frac{103380}{148157}=0.6977733080$ & $\frac{516902}{740785}=0.6977760079$ & $\frac{2}{740785}=2.6998\times10^{-6}$ \\
12 & $\frac{9666545}{13853391}=0.6977746459$ & $\frac{28999636}{41560173}=0.6977746700$ & $\frac{1}{41560173}=2.4061 \times 10^{-8} $\\
14 & $\frac{1310545608}{1878178855}=0.6977746579$ & $\frac{9173819258}{13147251985}=0.6977746580$ & $\frac{2}{13147251985}=1.5212 \times 10^{-10} $\\
\hline
\end{tabular}
\caption{Rigorous upper and lower bounds $a_{1}^{(m_0),\pm}$, and the allowed window $\Delta a_1^{(m_0)}= a_{1}^{(m_0),+}-a_{1}^{(m_0),-}$, for the allowed values of the moment $a_1$, or first Fourier mode, in our toy model \eqref{eq:toy1 angular model} for a fixed value of the parameter $\beta=2$. These are obtained by demanding $|a_{m_0}|<1$ together with the total-derivative recursion $a_{n+1} = -2\beta^{-1} n a_n + a_{n-1}$, for increasing values of $m_0$.}\label{tab:a 2nd cheap boot}
\end{table}

Although this cheap bootstrap already leads to highly tight rigorous bounds, at least in our toy model, this approach does not fully use the expected exponential decay behavior of high Fourier modes, which we will come to next.

\subsection{A  bootstrap on a shoestring}
\label{subsec:the cheap bootstrap}

Since the direct bootstrap method based on solving an SDP does not directly have any information about the growth/decay of $a_n$ in the numerical implementation, we can ask what happens in our toy model \eqref{eq:toy1 angular model} when we set $a_n=0$ at some large value and backtrack from there to extract $a_1$. 
For concreteness, consider the model \eqref{eq:toy1 angular model} at a value of $\beta=2$. Denoting $a_1=t$, the total-derivative recursion implies that the first few elements of the $a_n$ sequence are given by
\begin{align}
\begin{aligned}
    a_1 &=t ~,\quad &a_2 &= 1- t~,\quad & a_3 &= -2+3t ~,\\
    a_4 &=7-10t ~,\quad &a_5 &= -30+43 t ~,\quad & a_6 &= 157-223t ~.
\end{aligned}
\end{align}
Naively, since $|a_6|<1$, we find that $t\simeq 157/223 \pm 1/223$. 
The window of allowed values is the rigorous bound we discussed previously in the  cheap bootstrap. 
The center value of this window, which is nothing but the result of setting $a_6=0$, is then a good estimate for the actual value of the moment within the allowed window, with the error roughly the number of digits of the coefficient of $t$ appearing in $a_n$. 
Clearly, we get good estimate values of $t$ without having to solve a full semi-definite program. 
Let now check a higher order Fourier mode,  $a_{20}= 183586751854827751 - 263103209266016890 t$. Naively, we expect to be able to compute $t$ to about 18 digits of precision, but in practice we get twice as many digits correctly from simply setting $a_{20}=0$. Essentially setting $a_{20}\simeq \pm 1$ is too large an error bar for a quantity that should be exponentially suppressed. 

We will call this idea the \emph{shoestring bootstrap} method, as it is really cheap to find the solution. 
We simply estimate the true value of $a_1$ by setting $a_{n}=0$ for a sufficiently large $n$ and, using the recursion equations, solve for the corresponding value of $a_1$. This value, which we denote by $\bar{a}^{(n)}_1$, is an accurate estimate for the true value of $a_1$. 
This very simple procedure does \emph{not} give a rigorous two-sided bound (although a crude one is obtained from $|a_n|\leq1$). However, we may check whether the moment matrix $\mathcal{M}^{(k)}$ truncated to size $k$, evaluated for $a_1=\bar{a}^{({n})}_1$, is positive semi-definite. 
If true, this tells us that our estimate lies within the allowed region that would result from computing the rigorous two-sided bounds with the full SDP at level $k$.  
In other words, with the shoestring bootstrap we obtain a very accurate estimate at a very low computational cost (simply solving a linear equation), and may obtain a \emph{certificate of validity} by checking the PSD property of $\mathcal{M}^{(k)}$ once, without solving the full SDP.\footnote{For our toy model \eqref{eq:toy1 angular model}, $\bar{a}^{({n})}_1$ lies within the window implied by $|a_n|\leq1$, which in turn is a special case of the positivity of $\mathcal{M}^{(n+1)}$ as noted previously. Thus, the certificate of validity at level $n+1$ is automatic.} 

More generally, there will be a vector of undetermined Fourier modes (moments) $\vec t $. 
If $\dim(\vec t)=s$, since the map from $\vec t$ to the $a_{n+1} \dots a_{n+s}$ is affine (linear in $\vec t$ plus a shift), generically setting the $s$ values $a_{n+1}\dots a_{n+s}$ to zero determines a set of values for the components  of $\vec t$. This is the good guess at level $n$.
We need to solve a linear problem once for each level and check positivity of the matrix $\mathcal{M}^{(n+s)}$ only once to obtain a certificate of validity.
A naive error bar, if desired, may be computed from successive approximations to the variables that appear in $\vec t$ by changing the order at which the Fourier modes are set to vanish. 

Another way to justify this method is as follows. The recursion $a_{n+1}= a_{n-1}-2n\beta ^{-1}a_{n}$ for our toy model can be rewritten in terms of a matrix $S_n$ as follows
\begin{equation}\label{eq:matrix_rec}
    \vec v_n=\begin{pmatrix}
    a_{n}\\
    a_{n+1}
    \end{pmatrix}= \begin{pmatrix}
    0&1\\
    1&-2n\beta^{-1}
    \end{pmatrix}\begin{pmatrix}
    a_{n-1}\\
    a_{n}
    \end{pmatrix}=S_n \vec v_{n-1} ~.
\end{equation}
Notice that $|\det(S_n)|=1$. Starting from $\vec v_0$ to $\vec v_n$, the map is linear and of determinant equal to one in absolute value (and in this case the matrices are real).
For large $n$, the largest eigenvalue of $S_n$ is of order $-2n\beta^{-1}$ and it is large. Therefore there is also a small eigenvalue. In order for the sequence to be decreasing asymptotically, we expect that $|\vec v_{n+1}|< |\vec v_n| $, so the vector $\vec v_n$ needs to be dominated by the smallest eigenvalue. This has a very small $a_n$ relative to $a_{n-1}$. If we do the inverse map $\vec v_{n-1}= S_n^{-1} \vec v_{n}$, 
the vector that dominates is the largest eigenvalue of $S_n^{-1}$. This eigenvector dominates even if we set $a_{n+1}=0$, so that the corrections to $\vec v_{n-1}$ are small even with this slightly wrong vanishing value of $a_{n+1}$.
When we back propagate all the way to $\vec v_0$, $\vec v_0$ is proportional to the largest eigenvector of $(\prod S_k)^{-1}$. Since $a_0=1$, this determines $a_1$ to exponentially high accuracy.

A similar argument can be used for more general periodic potentials, where the determinant of the matrix
is of unit norm. This allows for the possibility of potentials that also depend on $\sin (k\theta)= [\exp(i k\theta)-\exp(-i k\theta)]/(2i)$ which introduce phases in the Fourier coefficient recursion relations.

In more general cases with more than one undetermined Fourier mode, one needs to work harder as $S_n$ becomes a $2\ell\times 2\ell $ matrix for some $\ell$, so there might be more than one small eigenvalue. Likewise, in the initialization of the sequence there are $\ell$ undetermined Fourier modes.
To illustrate this, we can consider a more nontrivial trigonometric measure of the form,
\begin{align}\label{eq:toy integral circle 2}
\int_{-\pi}^\pi d\theta\, \e^{\beta_1 \cos(m_1 \theta) + \beta_2 \cos(m_2 \theta)} ~,
\end{align}
where $m_1,m_2\in\mathbb{Z}_{>0}$ and without loss of generality we take $m_2>m_1$.
In this case, the total-derivative recursion relations for the Fourier modes take the form,
\begin{align}\label{eq:toy2 tot deriv relation}
a_n = a_{n-2m_2}-\frac{1}{m_2\beta_2}\left[ 2(n-m_2)a_{n-m_2} + m_1\beta_1 \left( a_{n-(m_2-m_1)} - a_{n-(m_2+m_1)} \right) \right]~.
\end{align}
This recursion relation determines all higher moments from the first $m_2$ moments, $a_1,\ldots,a_{m_2}$. That is, in this toy model we have $m_2$ independent variables: there are only $m_2$ unknowns, rather than $2m_2$. This makes the procedure well defined. The reason this works is that the recursion can also be applied with $n<2m_2$ where we get Fourier modes for negative $n$. For these we use $a_{-n}=a_n^*$ and if these are real, they are determined by previous data. All we need is that $|n-2m_2|<n$, or equivalently $n>m_2$. 
In the model \eqref{eq:toy integral circle 2}, it suffices to set $a_{n_0+j}=0$ for $j=1,\ldots,m_2$ and for a sufficiently large integer $n_0$, which we may view as a truncation parameter.

Tables \ref{tab:cheap boot model 2} and \ref{tab:sdp boot model 2} show a comparison between the conventional bootstrap bounds obtained from solving an SDP and our shoestring bootstrap estimate for the free (undetermined) Fourier modes $a_1,\ldots,a_5$, for the model \eqref{eq:toy integral circle 2} with $\{m_1,m_2\}=\{3,5\}$ and $\{\beta_1,\beta_2\}=\{\frac{4}{3},\frac{\pi}{\e}\}$. 
While the conventional bootstrap bounds do converge quickly, they become computationally more expensive to calculate as we increase the truncation level $k$. On the other hand, with our shoestring bootstrap method we can estimate the undetermined moments at much higher precision while still obtaining a certificate of the positivity of $\mathcal{M}^{(k)}$ for much higher values of $k$ at very low cost.\footnote{Although we only need one instance, obtaining this certificate by checking PSD of $\mathcal{M}^{(k)}$ becomes increasingly numerically unstable for very high $k$. Again, if we use the special case of rational arithmetic this is not an issue.} 
Here, we should compare the cutoff $n_0$ with an SDP at truncation level $k=n_0+m_2+1$, since in both cases the highest Fourier mode involved is $a_{n_0+m_2}$.
This simple example gives us confidence that our shoestring bootstrap method might prove very useful in more realistic bootstrap problems of physical interest. 
Understanding all the systematics in this and more general classes of trigonometric integrals is beyond the scope of the present paper.

\begin{table}[h!]
\centering
\begin{tabular}{|c|c|c|c|c|c|}\hline
$n_0$ & $\bar{a}_1^{(n_0)}$ & $\bar{a}_2^{(n_0)}$ & $\bar{a}_3^{(n_0)}$ & $\bar{a}_4^{(n_0)}$ & $\bar{a}_5^{(n_0)}$ \\
\hline
$38$ & $0.090721943934$ & $0.276253611322$ & $0.552285963926$ & $0.041860321092$ & $0.498944195169$ \\
$40$ & $0.090721944124$ & $0.276253611457$ & $0.552285963998$ & $0.041860321241$ & $0.498944195211$ \\
$42$ & $0.090721944107$ & $0.276253611444$ & $0.552285963994$ & $0.041860321227$ & $0.498944195210$ \\
$44$ & $0.090721944103$ & $0.276253611441$ & $0.552285963992$ & $0.041860321223$ & $0.498944195209$ \\
\hline
\end{tabular}
\caption{Estimated values for the free, or undetermined, Fourier coefficients obtained from our shoestring bootstrap method for the model \eqref{eq:toy integral circle 2} with $\{m_1,m_2\}=\{3,5\}$ and $\{\beta_1,\beta_2\}=\{\frac{4}{3},\frac{\pi}{\e}\}$. Here, we set $a_{n_0+j}=0$ for $j=1,\ldots,m_2$ and for increasing values of the truncation parameter $n_0$.}\label{tab:cheap boot model 2}
\end{table}

\begin{table}[h!]
\centering
\begin{tabular}{|c||c|c||c|c||c|c|}\hline
$k$ & $a^{(k)}_{1,-}$ & $a^{(k)}_{1,+}$ & $a^{(k)}_{2,-}$ & $a^{(k)}_{2,+}$ & $a^{(k)}_{3,-}$ & $a^{(k)}_{3,+}$ \\
\hline
$36$ & $0.0896391$ & $0.0918059$ & $0.2754800$ & $0.2770288$ & $0.5518769$ & $0.5526945$ \\
$38$ & $0.0898452$ & $0.0915991$ & $0.2755979$ & $0.2769099$ & $0.5520127$ & $0.5525593$ \\
$40$ & $0.0905527$ & $0.0908911$ & $0.2761174$ & $0.2763898$ & $0.5522320$ & $0.5523399$ \\
$42$ & $0.0906012$ & $0.0908427$ & $0.2761675$ & $0.2763397$ & $0.5522421$ & $0.5523299$ \\
$44$ & $0.0906845$ & $0.0907594$ & $0.2762277$ & $0.2762795$ & $0.5522720$ & $0.5522999$ \\
\hline
\end{tabular} \\ \vspace{0.25cm}
\begin{tabular}{|c||c|c||c|c|}\hline
$k$ & $a^{(k)}_{4,-}$ & $a^{(k)}_{4,+}$ & $a^{(k)}_{5,-}$ & $a^{(k)}_{5,+}$ \\
\hline
$36$ & $0.0410540$ & $0.0426682$ & $0.4987422$ & $0.4991453$  \\
$38$ & $0.0411303$ & $0.0425909$ & $0.4988279$ & $0.4990605$  \\
$40$ & $0.0417152$ & $0.0420055$ & $0.4989218$ & $0.4989665$  \\
$42$ & $0.0417711$ & $0.0419496$ & $0.4989224$ & $0.4989660$  \\
$44$ & $0.0418315$ & $0.0418891$ & $0.4989357$ & $0.4989527$  \\
\hline
\end{tabular}
\caption{Two-sided bounds $a^{(k)}_{j,\pm}$, with $j=1,\ldots,m_2$, obtained from solving an SDP at increasing truncation level $k$ for the model \eqref{eq:toy integral circle 2} with $\{m_1,m_2\}=\{3,5\}$ and $\{\beta_1,\beta_2\}=\{\frac{4}{3},\frac{\pi}{\e}\}$. These bounds are the result of the minimization/maximization of the first mode $a_1$.}\label{tab:sdp boot model 2}
\end{table}

Let us end this section with a comment on normalized vs un-normalized moments. It is important to remember that the bootstrap assumes that $a_0=1$ so the results are normalized. Is there a way to recover the un-normalized value of the integral? The answer is yes and it does not require integrating in the parameter $\beta$ from a known value as one would do in certain Monte-Carlo simulations to get the free energy. 
Recall that the eigenvalues of the positive Toeplitz matrix of Fourier modes are supposed to coincide asymptotically with the normalized angular distribution $d\mu$ itself. Alternatively, we can construct an approximation of the measure by computing a truncated Fourier series and choosing various points for comparison. After all, we are computing the Fourier coefficients of $d\mu$ with these methods. It must be the case that the maximum of $d\mu$ should be associated with the maximum eigenvalue of $\mathcal{M}$ and we can estimate this way the proportionality constant between the un-normalized distribution and the normalized one. We show an example of this in appendix \ref{app:unnormalized}. This structure of the eigenvalues of $\mathcal{M}$ is important in the study of matrix models, where we will apply these techniques.

\section{An application to unitary matrix models}\label{sec:matrix}

When we start with a measure like those appearing in \eqref{eq:measure}, they can be readily converted to a matrix integral as follows
$d\mu \propto d^{N^2} M_{ij} \exp(-\Tr(V(M)))$, or $d\mu \propto d^{N^2} U_{ij} \exp(-\Tr(\tilde V(U))+\tilde{\bar{V}}(U^{-1}))$, where we have substituted $x$ by an $N\times N$ hermitian matrix and taken a trace. A similar operation can be done with distributions on the circle provided they are written as polynomials $\tilde V(z= \exp(i\theta))$ and their complex conjugates. 

\paragraph{Orthogonal polynomials.}
In the classic work \cite{Brezin:1977sv} is was shown that such measures could be reduced to eigenvalue integrals, and that furthermore the models are solvable by the method of orthogonal polynomials for $\exp(-V(x))$ \cite{Bessis:1979is,Itzykson:1979fi}. What is interesting for us is that the orthogonal polynomials of $\exp(-V(x))$ can be computed from the moments of the distribution which we called $a_n$. Since the bootstrap technique in principle computes the $a_n$ to high accuracy (this is what is used in the semi-definite problem after all), then the bootstrap method can be said to implicitly determine the orthogonal polynomials. 

It is well known that numerically this problems suffers from poor conditioning \cite{gautschi1982generating,gautschi1985orthogonal} (see also \cite{gautschi2004orthogonal}). This means that the coefficients of the $k^{\text{th}}$ polynomial are very sensitive to numerical noise
at large $k$. Since the method of orthogonal polynomials essentially amounts to an orthogonalization with the matrix of moments $\mathcal{M}$ as a quadratic form, the fact that this matrix has a large hierarchy in its entries as well as its eigenvalues makes it easy to understand that if there are some cancellations, they can be very finely tuned and subject to noise. In the previous section, we argued that this was the reason that the bootstrap method produced exponentially fast convergence. If the matrix $\mathcal{M}^{(k)}$ fails to be positive at some $k$ (let us say there is exactly one negative eigenvalue that appears exactly at level $k$), we will find that in the process of orthogonalization the matrix $\mathcal{M}$ in the basis of orthogonal polynomials is diagonal and must have the same signature as $\mathcal{M}$. That is, the $k^{\text{th}}$ orthogonal polynomial will fail to have positive norm. In other words, we suspect that the reason that the problem of building the orthogonal polynomials is ill-conditioned is closely related to the reason  we get fast convergence in the bootstrap. A standard cure for this problem is to use very high precision arithmetic from the beginning. However, many SDP solvers only work with double precision libraries, so it might not be as useful as one would imagine naively.

For unitary matrix models there is a similar method where one can use orthogonal polynomials \cite{Periwal:1990gf} (a modern review of the applications of orthogonal polynomials to matrix models computations can be found in \cite{Marino:2008ya} and \cite{Eynard:2015aea}). The conditioning problem is less severe because the eigenvalues of the Toeplitz matrix $M$ are better behaved: the matrix is bounded after all. Moreover, we showed that what we called the shoestring bootstrap method did not rely at all on semi-definite programming. Instead, it was a problem of linear algebra where one checks positivity only once (per matrix $M_k$). If the couplings are such that the recursion relations have rational numbers, we can use exact linear algebra with  rational numbers to do the work. We just need to show that this extends to the computation of the corresponding orthogonal polynomials as well.

The definition of the orthogonal polynomials is as follows. There is a family of holomorphic monic polynomials $P_k(z=\exp(i\theta))$ indexed by the non-negative integer $k\in {\mathbb Z}_+$ with $P_0=1$ and $P_k=z^k+ \dots$ is of degree $k$. The inner product is orthogonal with respect to $d\mu$ as follows
\begin{equation}\label{eq:orthogonality of Pn}
    \mathcal{N}\int d \theta \exp(\tilde V(z)+\bar{\tilde{V}}(\bar z)) P_\ell(\bar z) P_m(z)= h_m \delta_{\ell,m} ~,
\end{equation}
where on the circle $\bar z = z^{-1}$ can also be used. 
Here, $\mathcal{N}^{-1}=\int d\theta \exp(\tilde V(z)+\bar{\tilde{V}}(\bar z))$ is again a normalization factor that ensures we are working with a normalized probability measure. 
Notice that the inner product is determined by the matrix $\mathcal{M}$, as these polynomials are finite Fourier series. The integral is defining an inner product for holomorphic function as follows
\begin{equation}
\braket f g = \int d\mu \bar f (\bar z) g (z) ~.
\end{equation}

In this setup, $P_n(z)$  is orthogonal to all polynomials of lower degree. When $\tilde V$ only has real coefficients (the measure is invariant under $\theta\to -\theta$) these holomorphic polynomials  satisfy a simpler recursion relation of the form
\begin{equation}\label{eq:recursion_poly}
    z P_n(z)= P_{n+1}(z) +  s_n z^n P_n(z^{-1}) ~, 
\end{equation}
and the $s_n$ are real numbers. The polynomials are therefore polynomials with real coefficients.
If we determine the $s_n$ we have determined the polynomials recursively.
Notice that $P_{n+1}$ must be orthogonal not only to $P_n$ but also to $z^n P_n(z^{-1}) \equiv \tilde P_n(z)$ which is the polynomial with reversed coefficients. 
We easily obtain that
\begin{equation}\label{eq:recursion_poly}
s_n = \frac{\braket{ \tilde P_n(z)}{z P_n(z)}}{\braket{ \tilde P_n(z)}{ \tilde P_n(z)}} ~.
\end{equation}
The denominator is also equal to $\braket{ z^n P_n(z^{-1})}{ z^n P_n(z^{-1})}= h_n$ and is positive.

Now we can prove that if the parameter $\beta$ is rational then the polynomials $P_n$ will have rational coefficients. 
The point is that the solution of the equation $a_k(a_1)=0$ is built from multiplying the matrices $S_n$ with rational coefficients and it is a linear equation whose solution is a rational number. By this method, all the $a_n$ coefficients are rational. 
If we assume that up to $\ell$ the  $P_\ell$ have rational coefficients, then $\braket {P_\ell}{P_\ell} $ is a sum of rational numbers. One can also check that $\braket{ \tilde P_\ell(z)}{z P_\ell(z)}$ is a rational number since the coefficients of $zP_\ell$ and $ \tilde P_\ell(z)$ are also rational. We conclude that $P_{\ell+1}$ is a rational combination of rational polynomials, hence it is also rational. 
Since $P_0=1$ and it is rational, we are done. This extends to more general recursions for general potentials as long as all the coupling constant parameters are rational numbers. The advantage of rational numbers is that they are not subject to floating point error. This is an issue for constructing the orthogonal polynomials as we will show.

Let us consider the Gross-Witten-Wadia model \cite{Gross:1980he,Wadia:1980cp}, which in fact has precisely the measure of the toy model \eqref{eq:toy1 angular model} we studied previously. The one important element we need to consider is how we normalize the couplings (the parameter $\beta$ of our measure) at large $N$. The idea is that $\beta=2 N/\lambda$ where usually $\lambda$ (the large $N$ coupling constant) is held fixed. The theory has a phase transition at $\lambda =2$, or equivalently at $\beta=N$.

\begin{figure}[ht]
\centering
\includegraphics[width=0.525\linewidth]{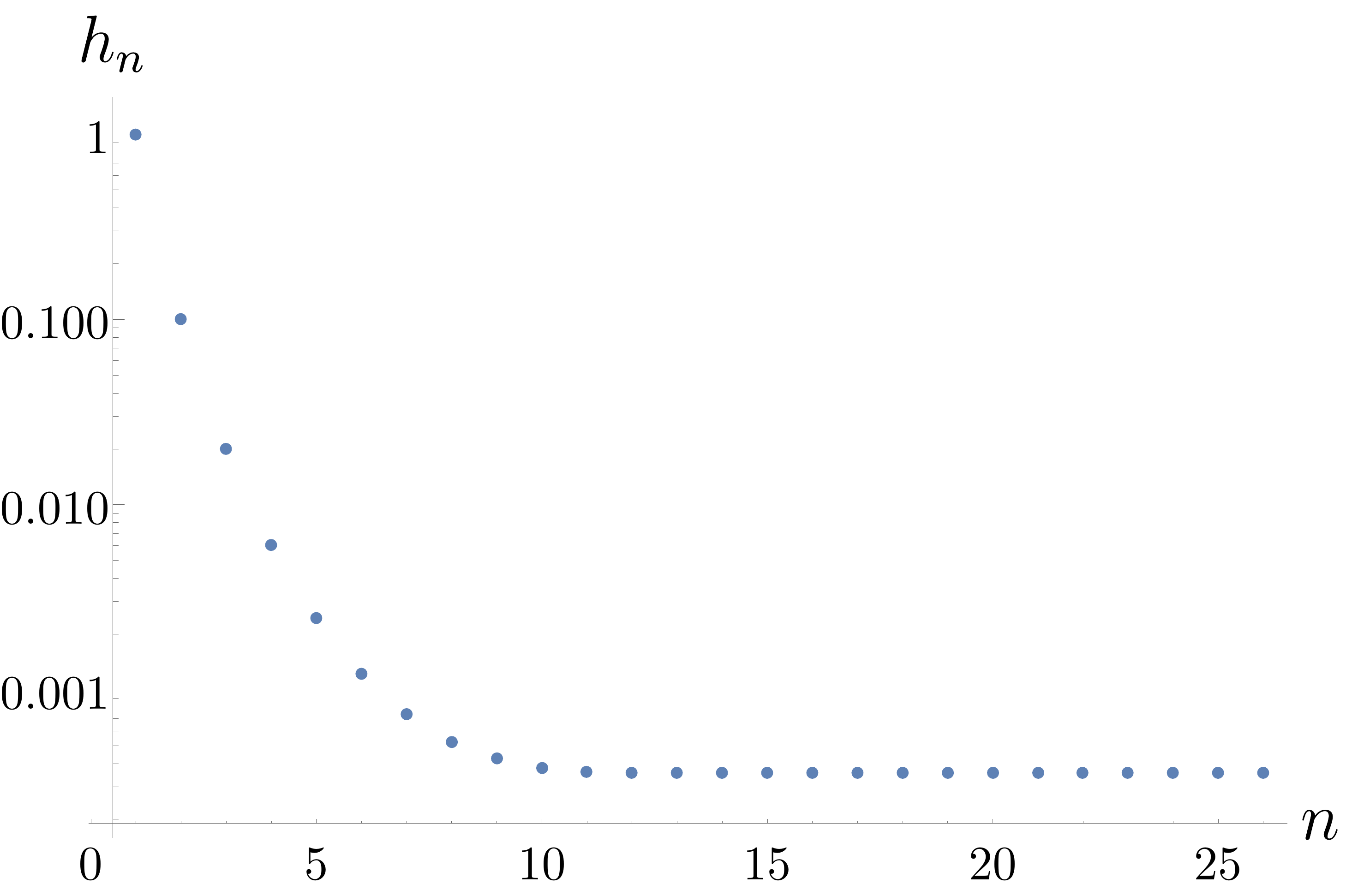}
\caption{Norms of the inner products $h_n=\braket{P_n}{P_n}$ for the case $\beta=10$ on a log scale.}
\label{fig:norms}
\end{figure}

The issue for us is the following. We want to study the theory at finite $N$, with $N$ large. Empirically, $N=10$ is usually close to infinity as the non-planar corrections are of order $1/N^2$, which one could call a $1\%$ error. 
The measure $d\mu = \exp(10 \cos(\theta))$ is sharply peaked around $\theta=0$. The maximum is $\exp(10)$ and the minimum is $\exp(-10)$.
As can be seen in appendix \ref{app:unnormalized}, this hierarchy  extends to the eigenvalues of the matrix of Fourier modes (the $\mathcal{M}^{(k)}$ matrices). What this means is that the matrix $\mathcal{M}$ is poorly conditioned (it has a large hierarchy in its eigenvalues). It is important to check the denominators in equation \eqref{eq:recursion_poly} and it turns out they tend to be small. This is shown in figure \ref{fig:norms}.

What we see is that even though the $P_{\ell}$ have coefficients of order one, their norm is very suppressed. This means that there are large cancellations at large $P_\ell$ and this produces numerical instabilities. In the present context, these only occur because of the large $N$ scaling. The large $N$ scaling is what forces $\mathcal{M}$ to have both very small and very large eigenvalues. In essence, the matrix $\mathcal{M}$ is close to being ill-conditioned. This means that we need the $a_m$ to be well known to very high precision to compensate for this problem numerically.

We can also look at our recursion relation and see how far we need to go to have very high precision.
Since in \eqref{eq:matrix_rec} we have a matrix of determinant minus one with real coefficients, there is a possibility that the eigenvalues of $S_n$ are very close in absolute value to one. If we want large eigenvalues in $S_n$, we need that $n\gg \beta \simeq N/\lambda$. That is, we need to solve the recursion at a value of $a_n=0$ where $n$ scales with $N$. The value of $n=N$ is the minimal value if we are building the orthogonal polynomials, because we need to know $a_{N-1}$ to build the $P_{N-1}$ orthogonal polynomial (recall we are counting from 0, so this corresponds to the first $N$ such polynomials). For instance, in the case $\beta=10$ we may take $n_0=200$ in our shoestring bootstrap method to determine $a_1$. Both the numerator and denominator in the resulting rational expression for $a_1$ are integers of order $10^{343}$. The value of $a_1$ computed with our shoestring bootstrap coincides with the normalized $a_1$ that can be computed analytically to better than 470 decimal places, and hence the associated orthogonal polynomials can be computed to very high precision as well. For example, the coefficients of the 25$^{\text{th}}$ orthogonal polynomial coincides with the one computed from the analytic result to about 460 digits of precision.

Notice now the following identity. Using equation \eqref{eq:recursion_poly}, we can take inner products and find that
\begin{equation}
\braket{zP_n}{z P_n} = \braket{P_{n+1}}{P_{n+1}}
+s_n^2\braket{\tilde P_n}{\tilde P_n} ~,
\end{equation}
or equivalently after a bit of algebra $h_{n+1}/h_n+s_n^2=1$. Notice that in figure \ref{fig:norms} the norms appear to become constant. This means that $s_n\to 0$ and $P_{n+1}(z)\simeq z P_n(z)$ for large $n$. This suggests that the coefficients of $P_n$ get shifted by one step each multiplication. We can say that the polynomials $\tilde P_n(z) = z^n P_n(1/z)$, which is $P_n$
with the coefficients reversed seems to converge in the large $n$ limit. Let us call this Taylor series $\tilde P_\infty(z)$.
What could this be? 

We can check that this implies that $\braket{P_n}{z^k P_n}\simeq 0$ for all small $k $ when $n$ is sufficiently large, since this is roughly $\braket{P_n}{ P_{n+k}}$ which vanishes identically. This is equal in the limit to 
$\braket{\tilde P_\infty}{z^k \tilde P_\infty}=0$ and these are Fourier coefficients of the form $\int  d\mu \exp(i k\theta)|P_\infty(z)|^2$.
If all of these vanish, except for $k=0$, we must conclude that $|\tilde P_\infty(z)|^2\exp( \tilde V(z) +\tilde V(1/z))$ is a constant. It is easy to check that this can be accomplished if $\tilde P_\infty(z) = \exp(-\tilde V(z))$. This also has the correct initial term $\tilde P_\infty(0)=1$. That is, $P_\infty(z)$ is the holomorphic inverse of the measure. Since for the bootstrap method we are using a normalized measure $d\mu(\theta)_{\text{Normalized}} = A d\mu(\theta)$, we have that $ \braket{\tilde P_\infty}{\tilde  P_\infty}= A \int 1  d\theta = 2\pi A$. We find then that we can compute $A$ analytically as $1/I_0(10)$ (see appendix \ref{app:unnormalized}), or equivalently $I_{0}(10)\simeq 1/\braket{\tilde P_\infty}{ \tilde P_\infty}$. We check this numerically by using $k=25$ to find that $I_0(10) $ and $1/ \braket{P_{25}}{ P_{25}}$ differ in the fifteenth digit. Essentially the convergence must be exponentially fast.

\begin{figure}[ht]
\centering
\includegraphics[width=0.55\linewidth]{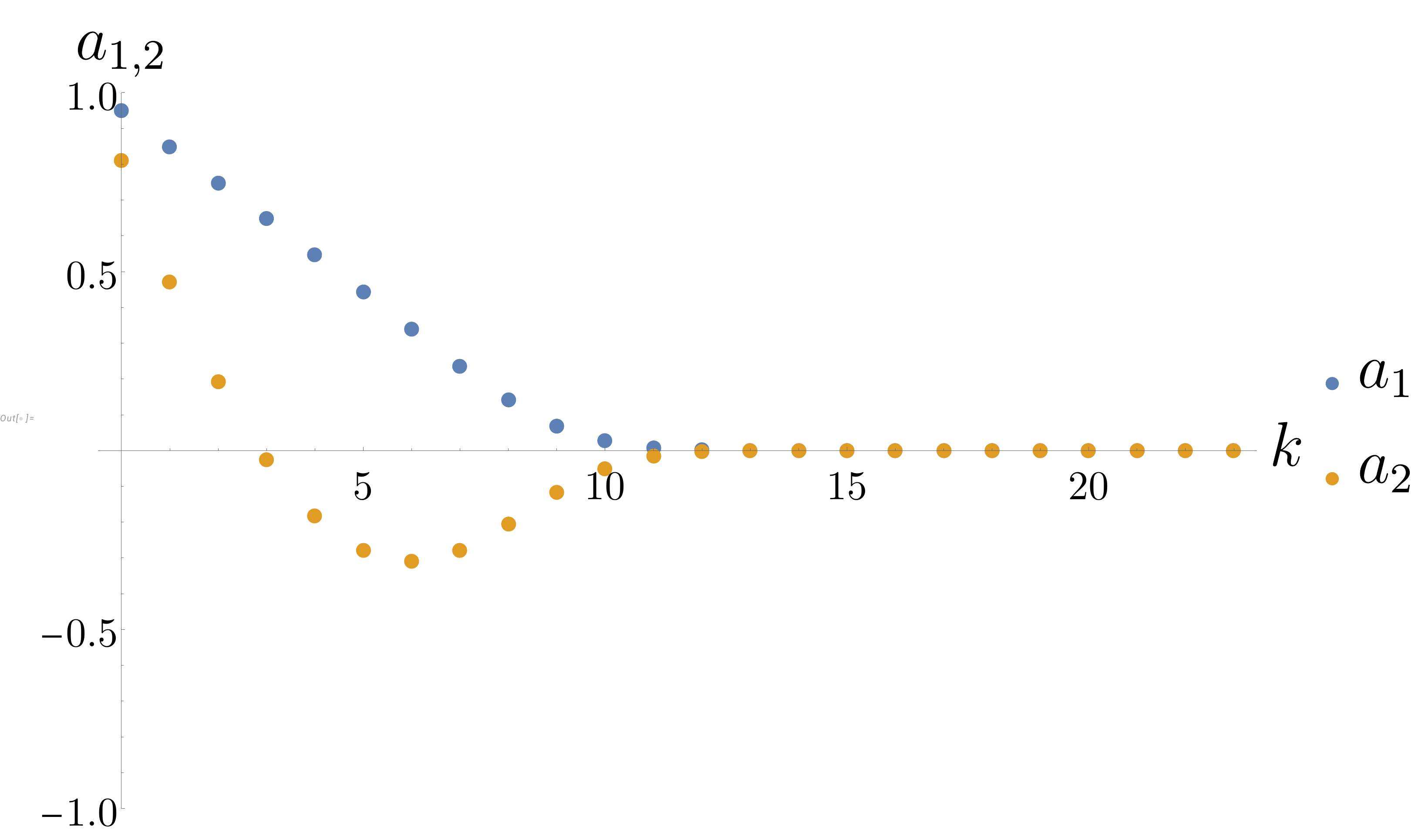}
\caption{First Fourier modes $a_{1,2}=\braket{P_k}{z^{1,2} P_k}/h_k $ for the case $\beta=10$ as a function of $k$.}
\label{fig:Fourier_poly}
\end{figure}

\paragraph{Wilson loops.}
Armed with the knowledge that the orthogonal polynomials seem to be precise enough, we can now study the contributions of each of these to the Wilson loop operators.
The idea is that in the matrix model one can reduce the problem to eigenvalues, so that the matrix integrals is given by
\begin{equation}
    d\mu \propto d^{N} \theta_j  \Delta(z_j) \Delta(z_j^{-1})\exp(-\sum\tilde{V}(z_j))+\bar{\tilde{V}}(z_j^{-1})) ~,
\end{equation}
where $\Delta(z_j)$ is the Vandermonde determinant of the $z$, given by matrix components $M_j^k= z_j^k$, with $k$ starting at zero and therefore 
\begin{equation}
    \Delta(z_j) = \det( z_j^k) ~.
\end{equation}
It is well-known that this determinant can  be written in terms of any basis of monic polynomials of the $z$ where 
\begin{equation}
    \Delta(z_j) = \det( P_k(z_j)) ~,
\end{equation}
and in particular, we can use the orthogonal polynomials we computed numerically. Because of orthogonality with respect to the exponential weight, it is straightforward to write the partition function in terms of the $h_i$. To compute the expectation values of the Wilson loops $W_n=\Tr(U^n)/N$, we need to write
\begin{equation}
W_n= \frac 1N \sum_{k=0}^{N-1} \frac{\braket{P_k}{z^n P_k}}{h_k} ~.
\end{equation}
These are essentially the sums of normalized Fourier coefficients of  the measures $d\mu P_k^*P_k $. We can analyze the terms individually and we can also understand that $W_n$ are the Fourier modes of the eigenvalue density on the circle. It is instructive to look at the values of $a_{1,2}=\braket{P_k}{z^{1,2} P_k}/h_k$ for increasing $k$ and at fixed $\beta=10$ as shown in figure \ref{fig:Fourier_poly}.

What is interesting is that both of these get close to zero and stay close to zero after $k=10$, so we can say that we are close to the $\tilde P_\infty$. Also the behavior of $a_1$ is linear between $0,10$. When we average the first few there is a distinct transition at $k=10$, corresponding to $\lambda=2$. This is the Gross-Witten transition. 

When we average over the first $N$ polynomials, the behavior changes at $N>10$. The reason is that the higher polynomials stop contributing. The result is the area under the linear curve, divided by $N$. Restoring the strong coupling constant $\lambda=\beta N$, the result behaves as $1/\lambda$. In the other region the average is linear in $N$, which gives a linear function of $\lambda$. This is continuous at $N=10$, or equivalently $\lambda=2$, exactly as the large $N$ Gross-Witten transition predicts. 

\begin{figure}[ht]
\centering
\includegraphics[width=0.55\linewidth]{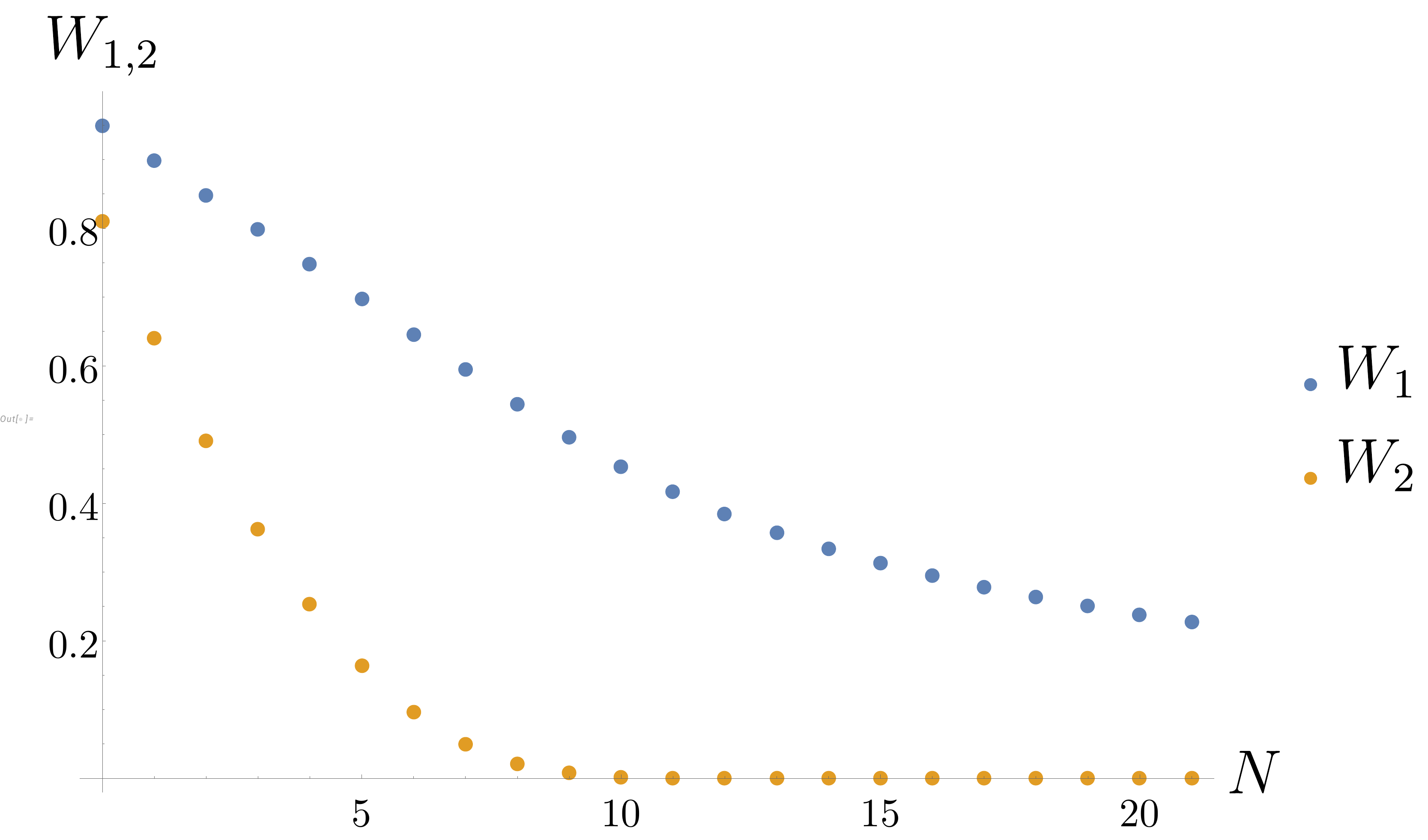}
\caption{Wilson loops $W_1$ and $W_2$ as a function of $N$ at fixed $\beta=10$.}
\label{fig:Wilson}
\end{figure}

\begin{figure}[ht]
\centering
\includegraphics[width=0.6\linewidth]{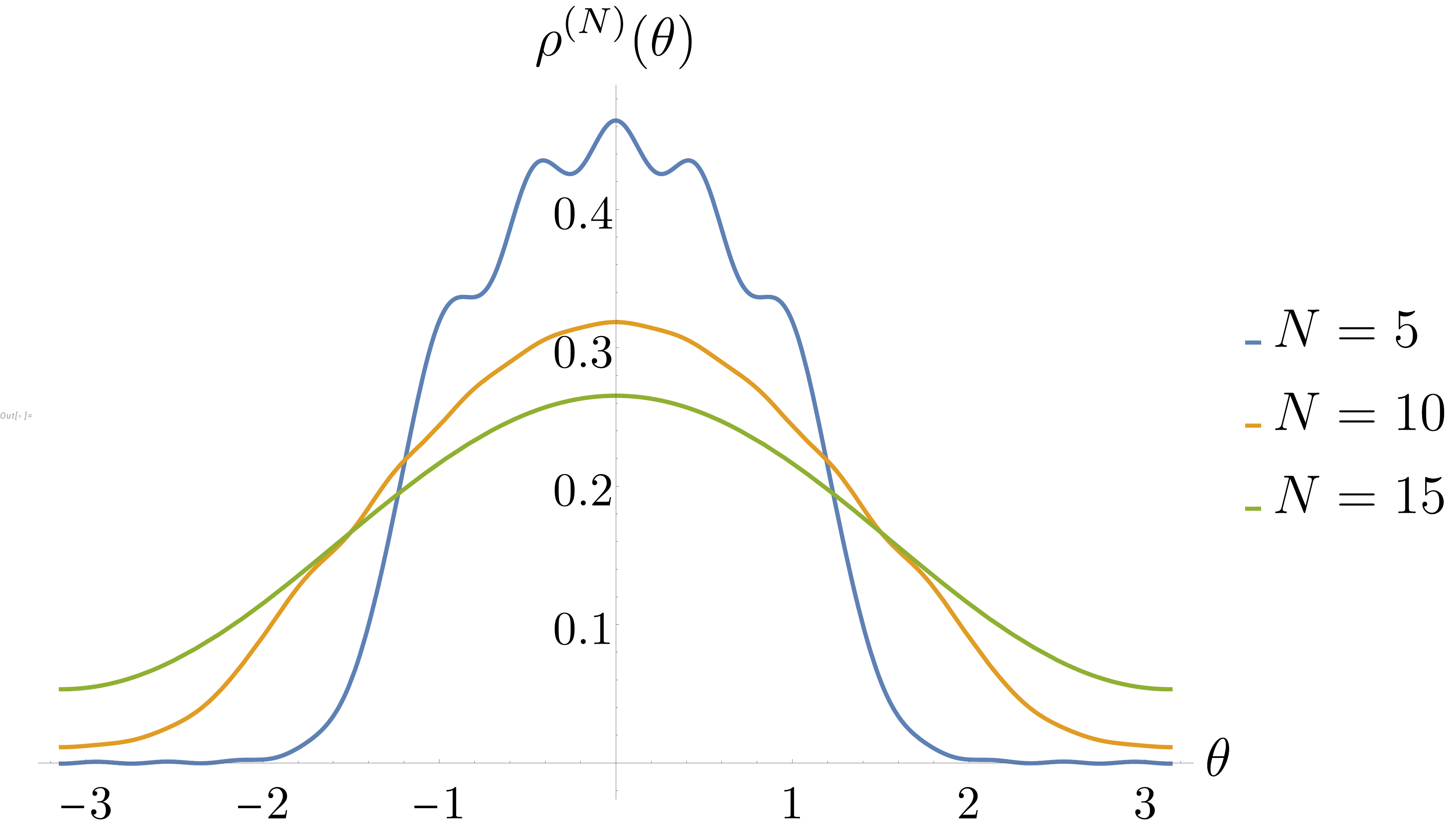}
\caption{Eigenvalue distribution density $\rho^{(N)}(\theta)$ at fixed $\beta=10$, for $N=5,10,15$.}
\label{fig:density}
\end{figure}

We can also see that $W_2$ vanishes numerically to a very good approximation beyond $N=10$ in figure \ref{fig:Wilson}, as predicted by the eigenvalue distribution of Gross-Witten, and that the transition is clear in $W_2$ at $N=10$.
For $W_1$, it looks smoother, as expected from the fact that this is a third order transition. Obviously, the difference with large $N$ is that this is the exact finite $N$ answer. 
The eigenvalue distribution is not quite uniform and can be computed from the Fourier transform of the $W_n$ (or directly from the $P_k(z)$) as $\rho^{(N)}(\theta)=\frac{1}{2\pi}\sum_{k\in\mathbb{Z}} W_k^{(N)}\e^{ik\theta}$. 
The result is depicted in \ref{fig:density}. Notice that for small $N$ the eigenvalue peaks are clear, so we can count that $N=5$.

It is instructive to compare these results to the bootstrap method for unitary matrix models described in \cite{Anderson:2016rcw}. In that case it is the loop equations of the matrix model that are studied and factorization is assumed. Indeed, this model is exactly solvable when studying the loop equations directly \cite{Friedan:1980tu}, so its study by positivity methods was a test of the method. What is important to understand is that it is the large $N$ solution that is easily found, not the finite $N$  solution. In the case where factorization is used, the loop equations are non-linear and not as easy to put in an SDP solver.  Anderson and Kruczenski sidestep this problem by minimizing an effective action instead and requiring a positivity constraint, finding good agreement.

There is another approximation scheme due to Marchesini and Onofri \cite{MARCHESINI1985225} which truncates the loop equations by setting some large $W_k=0$ and iteratively solves the problem. 
Alternatively, $W_k=0$ can be written as a polynomial equation in $W_1$ and one can find the root of those equations to find an approximation to $W_1$, as discussed in \cite{Anderson:2016rcw}. This constraint is somewhat similar to our shoestring bootstrap method, but the reasoning behind the constraint $W_k= 0$ is at least on its face different than ours.
Moreover, their method, which is nonlinear, proceeds by leveraging the large-$N$ loop equations. In contrast, our approach is linear, does not assume factorization, and can be applied directly at finite $N$.

 For more general unitary matrix models associated to pure gauge theory on the lattice, one can look at the results in \cite{Kazakov:2022xuh} for infinite $N$, which put the loop equations in SDP form by using convex relaxation.  For finite $N$, additional equations appear that reduce the number of possible loops
\cite{Kazakov:2024ool, Guo:2025fii} and the process gives good numerical results. We expect that constraints similar to our shoestring bootstrap can be leveraged so that the numerics in these approaches can be carried out faster and with far fewer computational resources.

\section{Conclusion}\label{sec:conclusion}

In this paper, we studied simple toy models of probability distributions on the real line and on the circle in order to better understand the fast convergence of certain typical  bootstrap problems based on loop equations, positivity, and convex optimization. 

For problems on the real line, we observed that the rapid convergence of the bootstrap method is linked to the behavior of the exact solution’s positive moment matrix: as the matrix size increases, its smallest eigenvalue decays exponentially, even as the moments themselves grow factorially.
Similarly, for problems on the circle, we observed that while the exact total-derivative recursion relation has factorially growing coefficients, the necessary asymptotic exponential decay of the Fourier modes of the distribution leads to a finely tuned positive moment matrix.
Both cases exemplify a Goldilocks phenomenon: the positivity condition of the associated moment matrix exhibits acute sensitivity to the initial conditions in the space of undetermined moments of the distribution. We find that this sensitivity is at the root of the fast convergence of the bootstrap method. 

For trigonometric integrals, this acute sensitivity led us to consider a much cheaper approach --- the \emph{shoestring bootstrap} method. Following the expected exponential decay of large Fourier modes (or moments), this method simply consists of setting a finite number of high Fourier modes to zero and using the exact total-derivative recursion relations to determine the remaining low Fourier modes. In particular, this requires solving a linear system of equations, which is much cheaper than the full bootstrap problem. We find that this method works remarkably well getting roughly double the precision of the full semi-definite program bounds and, at least in the example we studied, it is well justified analytically. We expect this idea to be useful in other contexts.

In contrast with the conventional bootstrap method based on solving a semi-definite program, which generally yields rigorous two-sided bounds on the moments of the distribution, the shoestring bootstrap method does not provide a rigorous bound but instead offers a very precise estimate of the moments.  
It is then inexpensive to check whether the positivity condition is satisfied, thereby obtaining a certificate of validity at a given truncation level: the shoestring program finds a point in the interior of the positive constraint region.
In other words, we can still be sure that our estimate lies within the rigorous bounds that a bootstrap SDP run would yield at that truncation level. 
With the shoestring bootstrap we can go well beyond the truncation level that would be computationally feasible for a full bootstrap SDP run.

A direct  application of our shoestring bootstrap method is to the study of unitary single matrix integrals and particularly to the Gross-Witten-Wadia model.
More specifically, we applied our method to accurately and numerically construct the  orthogonal polynomials associated to the matrix model, in terms of which observables of the matrix integral can be expressed nonperturbatively. 
Our method is simple enough that it can be implemented using rational or symbolic arithmetic, since it requires only solving a linear problem. 
This, in turn, allows us to compute the orthogonal polynomials with arbitrary precision, in contrast to commonly used SDP solvers that are limited to double or quad-double precision.
We also saw that constructing the orthogonal polynomials from the one eigenvalue measure is numerically unstable at large $N$, due to the poor conditioning of the positive matrix. Getting to very high precision  cures this instability. This high precision data is exactly the output of the shoestring bootstrap method.

Our main motivation for this study is the potential application of these techniques to lattice gauge theory in three and four dimensions, and to possibly overcome some of the bottlenecks that currently prevent large-scale, high-precision SDP bootstrap analyses for both finite and large $N$. We hope to report on this in the near future.

\acknowledgments
We would like to thank David Gross, George Hulsey, Clifford V. Johnson, P.N. Thomas Lloyd, and Mykhaylo Usatyuk for fruitful discussions and comments. D.B. work  supported in part by the Department of Energy under grant DE-SC
0011702.
VAR is supported by a DeBenedictis Postdoctoral Fellowship and funds from UCSB.

\appendix 

\section{Positive matrices and convexity}
\label{sec:review of matrices and convexity}

Here we collect some basic facts about positive matrices (operators) that we use to make statements about convergence of the bootstrap problem in the main text. These will serve to characterize the mathematical properties of how the bootstrap converges for certain classes of problems related to moment problems. These will also serve to understand how to improve bootstrap algorithms and make them more efficient.

Let ${\cal H}$ be a Hilbert space. Let $A$ be a self-adjoint operator acting on ${\cal H}$. We say $A$ is positive semi-definite and denote it by $A\succeq 0$ if 
\begin{equation}
\langle a | A |a \rangle \geq 0 \label{eq:positive_characterization}
\end{equation}
for all states in the Hilbert space. Care must be taken if $A$ is unbounded. For many of the cases we consider in this paper a finite Hilbert space will suffice most of the time, so in the following we will restrict ourselves to that case. We will use the shorthand $A$ is positive to indicate $A\succeq 0$.

The following lemmas are useful.
\begin{itemize}
\item {The eigenvalues of $A$ are greater than or equal to zero:} Let $\ket \lambda$ be a normalized eigenstate of $A$ with eigenvalue $\lambda$, that is $A \ket \lambda= \lambda \ket \lambda$ and $\braket \lambda  \lambda =1$. Then because of positivity we evaluate $\lambda \lambda \braket \lambda \lambda = \bra \lambda A \ket \lambda    \geq 0$.
\item {If the eigenvalues of $A$ are all greater than or equal to zero, then $A$ is positive:} write any vector in the basis the diagonalizes $A$ $\ket a= \sum a_\lambda \ket \lambda$. The result shows that 
\eqref{eq:positive_characterization} is satisfied.
\item {The set of positive semi-definite matrices is a convex cone.} If $r>0$, then $rA$ is positive. This makes the set of positive matrices into a cone. Convexity requires that if $A,B\succeq 0$, then $t A+(1-t) B\succeq 0$ whenever $t\in [0,1]$. This follows straightforwardly by checking \eqref{eq:positive_characterization} and verifying that a sum of positive numbers is positive. 
\end{itemize}

We say $A\succeq B$ if $(A-B)\succeq 0$. Let us consider now two self-adjoint operators $A, B$ with $A\succeq B$. The following are true.
\begin{itemize}
\item {The smallest eigenvalue of $A$ is greater than or equal to the smallest eigenvalue of $B$.} The proof is straightforward. Consider the normalized eigenket of the minimal eigenvalue of $A$, $\ket {\lambda_{\text{min}}}$. Then $\bra {\lambda_{\text{min}}} A \ket{\lambda_{\text{min}}}= \lambda_{\text{min}} \geq \bra {\lambda_{\text{min}}} B \ket{\lambda_{\text{min}}}$. If we now write $\ket{\lambda_{\text{min}}}$ in terms of the eigenvectors of $B$ (we use the spectral decomposition of $B$) the result follows. Similarly, the largest eigenvalue of $A$ is larger than the largest eigenvalue of $B$.

\item We call $A'$ a submatrix of $A$ if $A'$ is given by $A$ restricted to a linear subspace of ${\cal H}$.
If $A$ is self-adjoint, it follows that the minimal eigenvalue of $A$ is smaller than the minimal eigenvalue of $A'$. The minimal eigenvalue is characterized by the minimal value that the following function on the sphere takes $\lambda_{\text{min}}= \min_{\braket \lambda \lambda =1} \bra \lambda A \ket \lambda$. If $A'$ is a submatrix of $A$, the sphere associated to $A'$ is a subset of the sphere of $A$ and apart from that $A'$ is essentially the same as $A$. Thus, the minimal on the smallest sphere is larger than the global minimum for $A$.
Similarly, the largest eigenvalue of $A$ is larger than the largest eigenvalue of $A'$. 
These two results are a special case of the eigenvalue interlacing theorem. 

\item The function that computes the minimal eigenvalue of $A$ is convex, that is $\lambda_{\text{min}}(A)=\min(\text{eival}(A))$ is such that $\lambda_{\text{min}}(t A+(1-t) B)\geq \lambda_{\text{min}}(t A)+ \lambda_{\text{min}} ((1-t) B)= t \lambda_{\text{min}}( A) + (1-t) \lambda_{\text{min}} ( B)$.
The idea of the proof is to use second order perturbation theory to show that $\partial_t^2 \lambda_{\text{min}}(t A+(1-t) B) \leq 0 $. This is the usual statement that the second order perturbation of the ground state energy of a Hamiltonian that is a linear combination of two pieces $H_0+ g H_1$ is negative. It is usually written as follows
\begin{equation}
    \Delta^{(2)} E_0 = \sum \frac{|V_{i0}|^2}{E_0-E_i} \leq 0
\end{equation}
so that $\partial_g^2 E_0(g) \leq 0$. 
\end{itemize}

\section{Eigenvalue distributions of Toeplitz matrices}\label{app:unnormalized}
We want to numerically check how accurate is the approximate eigenvalue evaluation appearing in \eqref{eq:approx eval of circulant mat}.
We will also use this evaluation to evaluate un-normalized partition functions from the eigenvalue distributions.

For this procedure, we need to work to improve the eigenvalue estimation first.
Again, we want to consider the distribution $d\mu \propto \exp(\beta \cos(\theta))$ as an example. In this case the matrix
\begin{equation}
    \mathcal{M}= \begin{pmatrix}
    1 & a_1 & a_2 &\dots\\
    a_{-1} & 1& a_1 & \ddots\\
    a_{-2} & a_{-1} & 1 & \ddots\\
    \ddots & \ddots &\ddots &\ddots
 \end{pmatrix}
\end{equation}
is hermitian with real coefficients.

We need to consider that although there is approximate translation invariance in the center of the matrix, the matrix does not have exact periodic symmetry. Instead
the problem is more similar to a translation invariant lattice problem in one dimension with boundary conditions in an interval. In this case they are  the two corners of the matrix. With this in mind, the idea that there is a quasimomentum that can be diagonalized is not correct. We should  expect instead that momentum $k$ gets reflected into momentum $-k$ with some scattering phase. Because the matrix is real and the eigenvalues are real, the eigenvectors are also real and should generically have an equal superposition of $k,-k$ with a real coefficient. Essentially, we need to consider $k,-k$ together. This doesn't change the eigenvalue particularly, but the system should have no degeneracy in the spectrum of $\mathcal{M}$. We should only consider the angles between $0,\pi$ so that we only count one $k$. We compare the eigenvalue distribution with a uniform grid of points in the half circle, including the endpoints.

\begin{figure}[ht]
\centering
\includegraphics[width=8cm]{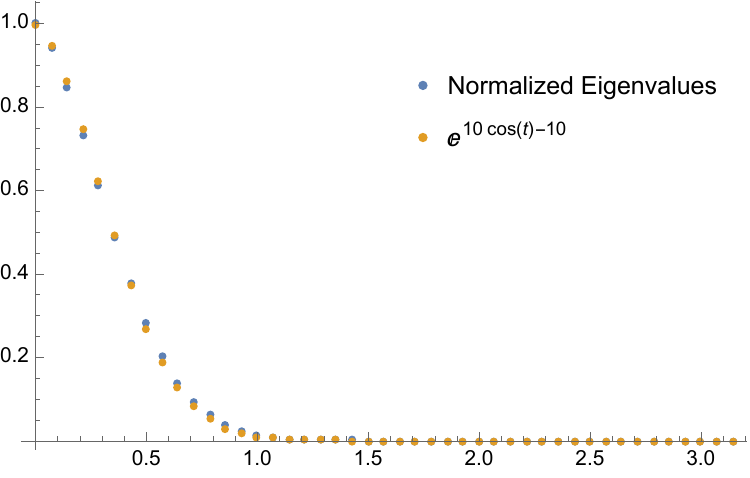}
\caption{Test of eigenvalue distribution in a $45\times 45$ matrix versus for $\beta=10$ and $d\mu(\theta)$. The eigenvalue distribution is normalized to the maximum eigenvalue and similarly the measure is normalized to the maximum.}
\label{fig:eival_compare}
\end{figure}

When we consider the example in figure \ref{fig:eival_compare}, we see broad agreement between the ordered eigenvalues and the distribution. Incidentally, because the finite matrix  $\mathcal{M}^{(k)}$  is a submatrix of $\mathcal{M}^{(k'>k)}$ we know that the maximal eigenvalue must be an increasing function of $k$. In the limit $k'\to \infty$ we must  be recovering exactly the maximum of the distribution, so even if we don't have complete information of the Fourier series of the distribution $d\mu(\theta)$ we get exact global bounds on the minimum and maximum of the distribution. 

We now have two distributions $d\mu(\theta)_N= A d\mu(\theta)$ where the second distribution is $d\mu(\theta)=\exp(\beta\cos(\theta))$ and where we have set $\beta=10$. Our goal is to estimate $A$ from the eigenvalue distribution. Of course, we can also approximate $d\mu(\theta)$ by the Fourier series and numerically maximize the result. This is not what we are doing here. We find that the maximum eigenvalue of the matrix is 
$\lambda_{\text{max}}\sim 7.6621118$, whereas $d\mu(\theta)_{\text{max}}=\exp(10)$. This suggests that $A\sim 7.6621118/ \exp(10) $.
Lastly, the expression \eqref{eq:approx eval of circulant mat} is not exactly the Fourier inverse; it differs by a normalization factor of $2\pi$ that needs to be taken into account.

Now we want to analyze the normalization of the total integral
\begin{equation}
    \int^{\pi}_{-\pi} d\theta\, \exp(\beta \cos\theta)= 2 \pi I_0(\beta) ~,
\end{equation}
and we need to compare to the zero-th Fourier mode of $d\mu(\theta)_N$. To take care of our normalization of Fourier coefficients, we need to divide by $2\pi$. Our estimate is then that
\begin{equation}
    I_0(10) =2815.7 \simeq 1/A= \exp(10)/7.6621118\sim2874.73 ~,
\end{equation}
so we find agreement to the first two decimal places. If we do the same estimate with matrices of size $65\times 65$, we get instead $I_0(10)\sim 2845.25$ which is better. 

  This is a proof of principle that even the un-normalized distribution (what one could call the partition function in statistical mechanics) can also be evaluated directly by the bootstrap ideas from eigenvalues, so long as we can compare to the un-normalized distribution somehow. Understanding the precise convergence of this method is beyond the scope of the present paper, as it would require a better understanding of the eigenvalue distribution for finite matrices, rather than the asymptotic large matrix limit.  
  This is an interesting problem in its own right.

\bibliographystyle{JHEP}
\bibliography{biblio.bib}

\end{document}